\documentclass[sigconf]{acmart}

\setcopyright{none}
\acmYear{2024}
\copyrightyear{2024}
\setcopyright{none}
\acmConference[CCS '24]{ACM Conference on Computer and Communications Security}{October 14-18, 2024}{Salt Lake City, USA}
\acmConference[CCS '24]{ACM Conference on Computer and Communications Security}{October 14-18, 2024}{Salt Lake City, USA}

\usepackage[vskip=1em,font=itshape,leftmargin=2em,rightmargin=2em]{quoting}
\usepackage[font=small,skip=0pt]{caption}

\usepackage[utf8]{inputenc}
\usepackage{subcaption}
\usepackage{fancyvrb}
\usepackage{multirow}
\usepackage{makecell}
\usepackage{array, multirow}
\usepackage{arydshln}
\newcommand{\ignore}[1]{}
\usepackage{balance}
\usepackage{xcolor}
\usepackage{booktabs}
\usepackage{adjustbox}
\usepackage{placeins}

\usepackage{wasysym}
\newcommand{\cmark}{\CIRCLE{}} 
\newcommand{\xmark}{\Circle{}} 

\newcolumntype{P}[1]{>{\centering\arraybackslash}p{#1}}
\usepackage{colortbl}
\usepackage{xcolor}
\usepackage{array}
\definecolor{Gray}{gray}{0.85}
\newcolumntype{H}{@{}>{\setbox0=\hbox\bgroup}c<{\egroup}}
\newcolumntype{a}{>{\columncolor{Gray}}c}
\newcommand{\rot}[1]{\rotatebox[origin=c]{90}{ ~#1~ }}  

\newcommand{\ccs}[1]{\textcolor{black}{#1}}
\newcommand{\mynote}[1]{\textcolor{orange}{[\textbf{NOTE:} #1]}}

\usepackage{epigraph}

\usepackage{tikz}
\newcommand*\circled[1]{\tikz[baseline=(char.base)]{
            \node[shape=circle,draw,inner sep=1pt,semithick] (char) {\small #1};}}

\newcommand{\status}[1]{\texttt{#1}}
\newcommand{\servfail}{\status{SERVFAIL}}

\newcommand{\must}{\textsf{MUST}}

\newcommand{\should}{\textsf{SHOULD}}

\newcommand{\del}[1]{}

\NewEnviron{delenv}{}

\begin{document}

\date{}

\title{The Harder You Try, The Harder You Fail:\\The KeyTrap Denial-of-Service Algorithmic Complexity Attacks on DNSSEC}
\author{Elias Heftrig, Haya Schulmann, Niklas Vogel, Michael Waidner}

\begin{abstract}
Availability is a major concern in the design of DNSSEC. To ensure availability, DNSSEC follows  Postel's Law [RFC1123]: \textit{"Be liberal in what you accept, and conservative in what you send."}
Hence, nameservers should send not just one matching key for a record set, but all the relevant cryptographic material, e.g., all the keys for all the ciphers that they support and all the corresponding signatures. This ensures that validation succeeds, and hence availability, even if some of the DNSSEC keys are misconfigured, incorrect or correspond to unsupported ciphers.

We show that this design of DNSSEC is flawed. Exploiting vulnerable recommendations in the DNSSEC standards, we develop a new class of DNSSEC-based algorithmic complexity attacks on DNS, we dub KeyTrap attacks. All popular DNS implementations and services are vulnerable. With just a single DNS packet, the KeyTrap attacks lead to a 2.000.000x spike in CPU instruction count in vulnerable DNS resolvers, stalling some for as long as 16 hours. This devastating effect prompted major DNS vendors to refer to KeyTrap as ``\textit{the worst attack on DNS ever discovered}''. Exploiting KeyTrap, an attacker could effectively disable Internet access in any system utilizing a DNSSEC-validating resolver.

We disclosed KeyTrap to vendors and operators on November 2, 2023, confidentially reporting the vulnerabilities to a closed group of DNS experts, operators and developers from the industry. Since then we have been working with all major vendors to mitigate \mbox{KeyTrap}, repeatedly discovering and assisting in closing weaknesses in proposed patches.
\ccs{Following our disclosure, the industry-wide umbrella CVE-2023-50387 has been assigned, covering the DNSSEC protocol vulnerabilities we present in this work.}
\end{abstract}

\maketitle

\section{Introduction}\label{sec:introduction}
The impact of the cryptographic requirements on the availability of DNS was a major concern in the design of DNSSEC [RFC4033-RFC4035]. Strict DNSSEC validation rules could impact DNS availability, hence, DNSSEC standard opted to limit strict requirements to the necessary minimum that suffices to ensure cryptographic security while maintaining availability of DNS, aiming at a trade-off between security, performance, and backward-compatibility. The standard requirements for DNSSEC were designed so that the DNS resolvers do not fail on the first cryptographic error. 
As long as a resolver can verify the provided information with any available DNSSEC material, the validation will succeed. 

 {\bf \textit{"Be liberal in what you accept, and conservative in what you send"} [RFC1123].} The core DNSSEC specification mandates validating DNS resolvers to try all possible keys when validating a resource record set (RRset) [RFC4035], and also strongly endorses to try all possible signatures covering it [RFC6840]. These DNSSEC requirements follow Postel's Law [RFC1123]: {\em the nameservers should send all the available cryptographic material, and the resolvers should use any of the cryptographic material they receive until the validation is successful}. This ensures availability even if some of the DNSSEC material cannot be used to validate authenticity of the DNS records, e.g., if the keys are misconfigured, incorrect or outdated. 
 We perform experimental evaluations and code analysis and find that these protocol requirements are supported by all major DNS resolver implementations.

{\bf DNSSEC algorithmic-complexity attacks.} In this work, we discover that the design philosophy of DNSSEC is flawed. We exploit the flaws in the DNSSEC standard and develop the first DNSSEC-based algorithmic complexity attacks against DNS. We demonstrate experimentally that our attacks are detrimental to availability of the affected DNS resolvers, leading to Denial of Service (DoS) on basic DNS functionalities, such as providing cached responses, or processing inbound or pending DNS packets. We show experimentally that an adversary using a single DNSSEC signed DNS response can DoS resolvers leading to a spike of 2.000.000x in CPU instruction count. The stalling period of the victim resolver depends on the resolver implementation, and can be up to 16 hours, see Table \ref{tab:hpocsq}. For comparison, a recently proposed NRDelegation attack \cite{afek2023nrdelegationattack} which exploited vulnerabilities in DNS to create multiple referral requests, would require 1569 DNS packets to cause a comparable increase in CPU instruction count, which our attacks achieve with a single packet. 
We find that all DNSSEC validating DNS software, DNS libraries and public DNS services on our dataset are vulnerable to our attacks; see list in Table \ref{tab:vulnerable:implementations:new}.

{\bf Flaws in DNSSEC.} We find that the flaws in DNSSEC specification are rooted in the interaction of a number of recommendations that in combination can be exploited as a powerful attack vector:

{\em Key tag collisions:} First, DNSSEC allows for multiple keys in a given DNS zone, for example during key rollover or for multi-algorithm support [RFC6781].  
Consequently, when validating DNSSEC, DNS resolvers are required to identify a suitable cryptographic key to use for signature verification.
DNSSEC uses key tag values to differentiate between the keys, even if they are of the same zone and use the same cryptographic algorithm. The triple of \textit{(zone name, algorithm, key tag)} is added to each respective signature to ensure efficiency in key-signature matching. When validating a signature, resolvers check the signature header and select the key with the matching triple for validation. However, the triple is not necessarily unique: multiple different DNS keys can have an identical triple. This can be explained by the calculation of the values in the triple. The algorithm identifier results directly from the cipher used to create the signature and is identical for all keys generated with a given algorithm. DNSSEC mandates all keys used for validating signatures in a zone to be identified by the zone name. Consequently, all DNSSEC keys that may be considered for validation trivially share the same name. Since the collisions in algorithm id and key name pairs are common, the key tag is calculated with a pseudo-random arithmetic function over the key bits to provide a means to distinguish same-algorithm, same-name keys. Using an arithmetic function instead of a manually chosen identifier eases distributed key management for multiple parties in the same DNS zone; instead of coordinating key tags to ensure uniqueness, the key tag is automatically calculated. 
However, the space of potential tags is limited by the 16 bits in the key tag field. Key tag collisions, while unlikely, can thus naturally occur in DNSSEC. This is explicitly stated in [RFC4034]\footnote{\path{https://datatracker.ietf.org/doc/html/rfc4035#section-5.3.1},\path{https://datatracker.ietf.org/doc/html/rfc4034#appendix-B}}, emphasizing that key tags are not unique identifiers. As we show, colliding key tags can be exploited to cause a resolver not to be able to identify a suitable key efficiently but to have to perform validations with all the available keys, inflicting computational effort during signature validation.

{\em Multiple keys:} Second, the DNSSEC specification mandates that a resolver must try all colliding keys until it finds a key that successfully validates the signature or all keys have been tried. This requirement is meant to ensure availability. Even if colliding keys occur, such that some keys may result in failed validation, the resolver has to try validating with all the keys until a key is found that results in a successful validation, ensuring the signed record remains valid and the corresponding resource therefore available. However, this "eager validation" can lead to heavy computational effort for the validating resolver, since the number of validations grows linearly with the amount of colliding keys. For example, if a signature has ten colliding keys, all with identical algorithm identifier, key tag and all invalid, the resolver must conduct ten signature validations before concluding the signature is invalid. While colliding keys are rare in real-world operation, we show that records with multiple colliding keys can be efficiently crafted by an adversary, imposing heavy computation on a victim resolver.

{\em Multiple signatures:} A philosophy, of trying all the cryptographic material available to ensure that the validation succeeds, also applies to the validation of signatures. Creating multiple signatures for a given DNS record can happen, e.g., during a key rollover. The DNS server adds a signature with the new key, while retaining the old signature to ensure the signature remains valid for all resolvers until the new key has propagated. Thus, parallel to the case of colliding keys, the RFCs specify that in the case of multiple signatures on the same record, a resolver must try all the signatures it received until it finds a valid signature or until all signatures have been tried. 

{\em ``The worst vulnerability ever found in DNS'':} We combine these requirements for the eager validation of signatures and of keys, along with the colliding key tags to develop powerful DNSSEC-based algorithmic complexity attacks on validating DNS resolvers. Our attacks allow a low-resource adversary to fully DoS a DNS resolver for up to 16h with a single DNS request. The task force with $31$ participants from the major operators, vendors and developers of DNS/DNSSEC, to which we disclosed our research, dubbed our attack: {\em the most devastating vulnerability ever found in DNSSEC}. 
{\bf Complex vulnerabilities are challenging to find.} Surprisingly the flaws are not recent. The requirement\footnote{\path{https://datatracker.ietf.org/doc/html/rfc2535#page-46}} to try all keys was present already in the obsoleted [RFC2535] from 1999. \ccs{This requirement, to try all the keys, was carried over to the current specification [RFC4035]. In 2013 the issue was further exacerbated by [RFC6840] recommending validators to also try all the signatures.}

The vulnerabilities have been in the wild since at least August 2000 in Bind9 DNS resolver\footnote{\path{https://github.com/isc-projects/bind9/commit/6f17d90364f01c3e81073a9ffb40b0093878c8e2}} and were introduced into the code\footnote{\url{https://github.com/NLnetLabs/unbound/commit/8f58908f45d69178f8a30125d8ebcedf3c6f6761}} of Unbound DNS resolver in August 2007. 
Using the code of Unbound as an example, the vulnerable code performs loops over keys and signatures:

{\scriptsize
\begin{verbatim}
//loop over all the keys                     //loop over all the signatures
for(i=0; i<num; i++) {                       for(i=0; i<num; i++) {
   /* see if key matches keytag and algo */      sec = dnskeyset_verify_rrset_sig
   if(algo != dnskey_get_algo(dnskey, i) ||          (env, ve, rrset, dnskey, i);
      tag != dnskey_calc_keytag(dnskey, i))      if(sec == sec_status_secure)
      continue;                                      return sec; 
   numchecked ++;                              }
   /* see if key verifies */
   sec = dnskey_verify_rrset_sig
        (env, ve, rrset, dnskey, i, sig_idx);
   if(sec == sec_status_secure)
        return sec;
}
\end{verbatim}
}

Although the vulnerabilities have existed in the standard for about 25 years and in the wild for 24 years, they have not been noticed by the community. This is not surprising since the complexity of the DNSSEC validation requirements made it challenging to identify the flaws. The exploit requires a combination of a number of requirements, which made it not trivial even for DNS experts to notice. The security community made similar experiences with much simpler vulnerabilities, such as Heartbleed or Log4j \cite{heartbleed,log4j} which were there but no one could see them, and they took years to notice and fix. Unfortunately, in contrast to these vulnerabilities, the vulnerabilities we find in this work are not simple to resolve, since they are fundamentally rooted in the design philosophy of DNSSEC, and are not just mere software implementation bugs. 

{\bf Flaws are challenging to mitigate.} The flaws in DNSSEC validation are not simple to solve. There are legitimate situations in which nameservers may return multiple keys, e.g., to account for potential failures. For instance, some domains may be experimenting with new ciphers, not yet supported by all the resolvers and there are keys-rollover. To avoid failures the nameservers should return all the cryptographic material. Similarly, to ensure successful validation, the resolvers should not fail on the first unsuccessful validation, but should try all the material until validation succeeds. Indeed, the experience we made since we have started working on the patches with the developers shows that these flaws can be substantially mitigated, but cannot be completely solved. Attacks against the final patch still result in heavy CPU load caused by high attacker traffic, but at least DNS packet loss is prevented.
Solving these issues fundamentally requires to reconsider the basics of the design philosophy of the Internet. 

{\bf The importance of understanding and weaponizing vulnerabilities.} While there were concerns in the community that key tag collisions could introduce a weakness\footnote{\url{https://ripe78.ripe.net/presentations/5-20190520-RIPE-78-DNS-wg-Keytags.pdf}} and even a bachelor project attempted to find an attack\footnote{\url{https://essay.utwente.nl/78777/1/Research_paper.pdf}}, no compelling method to weaponize the key tag and demonstrate an attack was found. Therefore, the collisions were not regarded as a practical threat and vendors did not issue patches. Understanding how to exploit and weaponize a vulnerability and the ability to demonstrate it to the community is critical. There are numerous examples where the ability to understand and realize a threat led to improvements in the security landscape. One such example in DNS is that of port randomization. Initially the DNS resolvers were using predictable or fixed source ports for their requests, until a security expert Kaminsky found a way to weaponize and demonstrated a practical DNS cache poisoning attack \cite{kaminsky:dns}. Although predictable DNS source ports were seen by many experts in the community as a threat, this was not fixed until a practical attack was demonstrated\footnote{\url{https://makezine.com/article/technology/djbdns-dns-exploits-bernstein/}}. Prior to Kaminsky's demonstration, despite the concerns, the vendors did not consider this a practical threat. Following Kaminsky's attack, all the vendors quickly patched their DNS resolvers to send DNS requests from unpredictable ports. This and other examples show that to improve security a deep understanding of the problem at hand is required, in order to find a way to weaponize it, so that it then becomes an attack. This was also the case with key trap, after we found a way to demonstrate for the first time that key trap was a practical threat, all the vendors immediately issued patches.

{\bf Contributions.} We make the following contributions:

$\bullet$ Conceptually, we find that the aim to ensure validation at any cost in DNSSEC standard exposes the DNS resolvers to attacks. We analyze the DNSSEC standards in §\ref{sec:vulnerability}, and identify flaws in the DNSSEC standards which enable complexity attacks. 

$\bullet$ We combine the flaws in the RFCs to develop the first algorithmic complexity attacks that exploit vulnerabilities in DNSSEC. We find experimentally that all standard-compliant DNS implementations support the flawed recommendations and hence are vulnerable to our attacks.

$\bullet$ We analyze the code of popular DNS implementations to identify the effects of the stalling on, e.g., caching, pending DNS requests or inbound/pending DNS packets. We use our observations to provide recommendations for adapting the architecture of the resolvers to improve the robustness to failures and attacks.

$\bullet$ We performed ethical disclosure of our vulnerabilities to the major DNS vendors, DNS/CDN/cloud operators and standardizers on November 2, 2023. Since then, we have been intensively working with this group on developing patches and regularly communicating with the developers within a closed chat group. We provided to the developers our attack vectors encoded in DNS zonefiles and set up a test environment for evaluation of vulnerabilities in DNSSEC, which alleviates the need for manual setup and enables quick evaluation of the attacks against the proposed patches. We provide a timeline for disclosure and of the patches development process.
\ccs{Our discovered vulnerabilities were assigned an umbrella CVE in December 2023.}

{\bf Organization.} We compare our research to related work in §\ref{sec:related-work}. We provide an overview of DNS and DNSSEC in §\ref{sc:overview}. We analyze the recommendations in the DNS standard specification in §\ref{sec:vulnerability}. We construct the attacks in §\ref{sec:attack} and evaluate them against major DNS implementations and services in §\ref{sec:praceval}. Disclosure and the process of developing mitigations are in §\ref{sec:countermeasures}. We \ccs{discuss ethical considerations in §\ref{sec:ethics} and} conclude in §\ref{sec:conclusions}.

\section{Related Work}\label{sec:background-related-work}\label{sec:related-work}
In Distributed Denial of Service (DDoS) attacks adversaries flood a victim resource, e.g., a network bandwidth or a buffer, with a large volume of packets loading the target victim beyond its available capacity and causing packet loss \cite{kuhrer2014exit}. DNS is often a victim of DDoS, either as a target or as a reflector to flood other victims. Since DNS responses are larger than DNS requests, reflected DNS responses amplify the load generated by the attacker's requests. The amplification factor is exacerbated with DNSSEC, whose signatures and keys further increase the sizes of DNS responses \cite{van2014dnssec}. 
An amplification effect can also be achieved by exploiting vulnerabilities in protocol implementations or vulnerabilities in the processing of DNS records \cite{bushart2018dns,DBLP:conf/imc/MouraCHWH20}. 
NXNSAttack \cite{afek2020nxnsattack} exploited a vulnerability that generated a flood of queries between the recursive resolver and the authoritative server creating a load on them both. Recently, \cite{afek2023nrdelegationattack} demonstrated a complexity attack on DNS which causes a victim resolver to issue multiple queries, following delegation responses by a malicious authoritative server. The victim resolver issues the queries to nameservers which do not respond, eventually exhausting its resources. 

The NRDelegation attack in \cite{afek2023nrdelegationattack} is shown to achieve a 5600x increase in CPU instructions between the attack requests and benign DNS requests. In contrast, our KeyTrap attack in this work achieves a 2000000x increase in CPU instruction count. 

To compare the impact between both attacks on the CPU instruction count, we set up a benign and a malicious signed domains.
We set up an Unbound resolver in an isolated environment and run Linux perf to measure CPU instruction count. We first measure the CPU instruction count of a request to a benign DNSSEC signed domain.
To ensure reliability we average out the instruction count over five measurements. Further, we set up the attack domain on the same DNS server. The measurements are conducted on Ubuntu 22.04 LTS with Unbound 1.19.0 DNS software. In our test setup, we find that a benign request on a signed DNSSEC domain requires approx. \textbf{811.500 CPU instructions} on Unbound. In contrast, we find a significantly higher instruction count for the resolution of the KeyTrap attack domain. To resolve and validate the domain, Unbound takes approximately \textbf{1.725.000.000.000 CPU instructions}. Comparing to the benign request, the attack thus leads to a 2.000.000x increase in CPU instruction count, compared to the 5600x increase in NR delegation. Directly comparing the CPU instructions count of \cite{afek2023nrdelegationattack}, we find that NR delegation attack requires 1569 queries to result in the same increase in CPU instruction count as a single request with our KeyTrap attack. Hence, a KeyTrap request leads to the same load as approx. 2 million benign requests.

Van Rijswijk-Deij et al. \cite{van2016performance} explored the performance of ECC vs RSA on Bind9 and Unbound. 
They evaluated the load on the Bind9 and Unbound resolvers when sending multiple signatures and found that the ECC algorithms do not impose too much additional CPU load on the two resolvers in contrast to RSA. 
To create load the authors made the resolver request a large number of non-existent records (NSEC3), causing many DNS responses, each carrying up to three NSEC3 records plus one SOA, with one signature validation per record. In effect, the victim resolver was validating four RRSIG records per response.
While the responses sent by \cite{van2016performance} caused the resolver to perform 4 validations, equivalent to the number of signatures their nameserver returned (an order of $\mathcal{O}(n)$), our specially crafted records trigger more than 500K validations per DNS response (an order of $\mathcal{O}(n^2)$). Our attack scales quadratically with the number of keys returned to the resolver.

In contrast to previous work our KeyTrap attacks do not require multiple packets, instead we exploit algorithmic complexity vulnerabilities in the DNSSEC validation in DNS resolvers as a building block to develop CPU based DoS attacks. Our complexity attacks are triggered by feeding the DNS resolvers with specially crafted DNSSEC records, which are constructed in a way that exploits validation vulnerabilities in cryptographic validation logic. 
When the DNS resolvers attempt to validate the DNSSEC records they receive from our nameserver, they get stalled.
Our attacks are  extremely stealthy, being able to stall resolvers between 170 seconds and 16 hours (depending on the resolver software) with a single DNS response packet. All the resolvers we tested were found vulnerable to our attacks. We evaluate how DNS implementations react to the load created by the attack and find that certain design choices can enable faster recovery from our DoS attacks. 

Our work is also related to downgrade attacks against DNSSEC, DNSSEC \cite{heftrig2023downgrading}. The DNSSEC-downgrade attacks however focus on disabling DNSSEC validation, but do not have adverse effects on the availability of the victim resolvers.


\section{Overview of DNSSEC}\label{sc:overview}

\ignore{
\begin{figure}[t!]
    \centering
    \includegraphics[width=0.7\columnwidth]{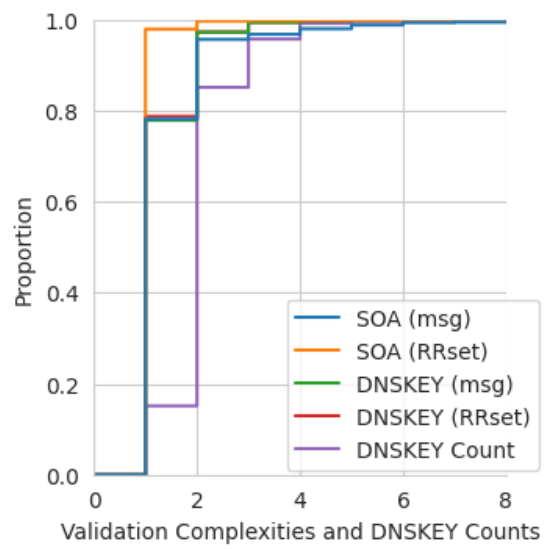}
    \caption{Validation complexities in responses from domains.\mynote{Ugly figure. To be reworked.}}
    \label{fig:rpki_fig}
\end{figure}
}

DNSSEC [RFC4033-4035] ensures origin authenticity and data integrity in DNS. To gain security benefits the domain owners should digitally sign the records in their domains, and should upgrade the nameservers to serve DNSSEC-signed DNS responses. The DNS resolvers should validate the received DNS records against the digital signatures. To validate the public keys the resolvers should construct a validation path from the root zone to the target domain. If validation fails, the resolver must not deliver the bogus records to the client and instead signal an error by sending a \servfail{} response. If DNSSEC validation is successful, the resolver should return the requested records to the client and cache them.


DNSSEC signatures are conveyed in RRSIG-type DNS records. An RRSIG record is associated with the set of records (RRset) it covers by name, class, and the record type indicated by an RRSIG-specific record field. 
The public keys used to validate the signatures are sent in DNSKEY-type records.

There are two types of keys: Zone-Signing-Key (ZSK) and Key-Signing-Key (KSK).
The ZSKs are used to sign records in the zone and are authenticated with a KSK.
\ccs{DNSKEY records contain multiple fields, including usage-indicating flags, the protocol and algorithm specifiers, and the key bytes.
From these record data the key tag can be calculated. The KSKs and all ZSKs of a zone are included into a DNSKEY set which is signed by at least one KSK. Signature records covering the DNSKEY set need to reference the key tag of a KSK. Only after the DNSKEY set is validated can the ZSK be used to validate signatures covering other records (RRs). To support simple DNSSEC setups, KSK and ZSK can be identical.}

DS records from a parent zone are used to authenticate individual KSK type DNSKEY records in a child zone. This is done to delegate trust from a parent zone public key to a child zone public key. DS records use the same triple \textit{(zone name, algorithm, key tag)} as RRSIGs to identify a subset of candidate DNSKEYs.

DNS records contain mappings from DNS names to resources. In this work, we use the DNS A record in the evaluations of our attacks. The A record contains the mapping of a domain name in the zone to an IPv4 address. The A record includes a TTL value, which specifies the validity period for caching. The A-type record is queried by a resolver when resolving a domain name. For instance, an A record may map the domain \textit{www-x.attack.er} to the IP address 6.6.6.6. We explain the functionality of DNS and DNSSEC with concrete examples in §\ref{sec:attack}.

\section{Analysis of DNSSEC Specification}\label{sec:vulnerability}
In the following, we illustrate the validation recommendations in the DNSSEC standard relevant to the KeyTrap attacks.

{\bf Associating keys with signatures.} A domain can use multiple different keys, see [RFC6840, §6.2]. This is required for instance to support new stronger cryptographic ciphers but also to offer weaker ciphers for other non-supporting resolvers, or to support stronger and weaker keys of same cipher, or during key rollover. 

In such a situation the DNS records are signed with all those keys and the resolver receives the keys and signatures in a DNS response. 

To authenticate an RRset the RRSIG covering it needs to be associated with the DNSKEY record carrying the corresponding public key. This is done by matching the \texttt{Signer's Name} in the RRSIG record data field with the name of the DNSKEY record and the algorithm fields. 
Additionally, the value of the \texttt{Key Tag} field in the RRSIG must match the key tag value of the DNSKEY.
Note that the DNSKEY record data format does not specify a \texttt{Key Tag} field. Instead, the key tag value is calculated by resolvers as an unsigned 16-bit integer sum over all two-octet words in the DNSKEY record data (ignoring carry and allowing overflow).
As highlighted by [RFC4034, §B], the key tag is not a unique identifier, but a mechanism to efficiently identify a subset of DNSKEYs possibly matching the RRSIG to be validated.
In consequence, to successfully authenticate an RRset covered by an RRSIG, the resolver \must{} try all DNSKEYs in this subset until it succeeds to validate the signature using one of the candidate public keys or runs out of keys to try [RFC4035, §5.3.1].

Moreover, the DNSSEC key tag is not a crytographic fingerprint of the public key. 
Representing an unsigned integer sum over the record data the key tag does not provide a cryptographic collision resistance. In §\ref{sec:attack} we develop an attack \textit{LockCram} which exploits the requirement to associate keys with signatures.

{\bf Resolvers are endorsed to try all signatures.}
To support a variety of domain-side key and algorithm roll-over schemes, as well as to increase robustness against cache-induced inconsistencies in the Internet-wide DNS state, resolvers must be tolerant in case individual signatures do not validate.
Besides ignoring any RRSIGs, which do not match any authenticated DNSKEY, resolvers are endorsed by specification (\should{}) to try all RRSIGs covering an RRset until a valid one is found. Only if all signatures fail to authenticate the RRset should the resolver mark it invalid. When multiple RRSIGs cover a given RRset, [RFC6840, §5.4] suggests that a resolver \should{} accept any valid RRSIG as sufficient, and only determine that an RRset is bogus if all RRSIGs fail validation.

The explanation is that if a resolver adopts a more restrictive policy, there is a danger that properly signed data might unnecessarily fail validation. . 
Furthermore, certain zone management techniques, like the Double Signature Zone Signing Key Rollover method described in [RFC6781, §4.1.1], will not work reliably.

{\bf Resolvers try to authenticate all DNSKEYs with all DS hashes.} The DNSSEC standard is not clear on the requirement of DS hashes authentication. This vagueness left it for developers to decide how to implement the DS validation. We experimentally find that all the resolvers in our dataset validate all the DS hashes.

{\bf RFC-compliant resolvers are vulnerable.} We find experimentally that all the resolvers in our dataset adhere to RFC specifications, validating all signatures with all DNSSEC keys they received from the attack server and validate all DS hashes against all the DNSKEY records. For examples, see the the validation routines in Unbound\footnote{\url{https://github.com/NLnetLabs/unbound/blob/master/validator/val_sigcrypt.c}\\lines $641$ and $704$.}. 
In this work we show that these requirements are vulnerable. 
We develop KeyTrap algorithmic complexity attacks that exploit the specification weaknesses in the association process described above to forge a DNSKEY set of cardinality $k$, conforming to a single key tag $t_k$, and to create a large number $s$ of invalid RRSIG records, which all reference these DNSKEYs. In consequence, the resolver needs to check all $s$ signatures against all $k$ keys -- a procedure with asymptotic complexity in $\mathcal{O}(n^2)$. 

\section{Resource Exhaustion Attacks}\label{sec:attack}
Our attacks consist of a module for sending queries to the target resolver, malicious nameservers and the zonefiles that encode the KeyTrap attack vectors.
We exploit algorithmic complexity vulnerabilities in standard requirements to develop different variants of KeyTrap resource exhaustion attacks: KeySigTrap, SigJam, LockCram, and HashTrap. To initiate the attacks our adversary causes the victim resolver to look up a record in its malicious domain. The attacker's nameserver responds to the DNS queries with malicious record sets (RRsets), according to the specific attack vector and zone configuration.

\subsection{Threat Model}\label{sc:threat-model}
\ccs{In our work we consider a low-resource attacker who is capable of hosting a DNSSEC-signed domain with a secure delegation and who can attract a victim to resolve a name in this domain.
Hosting a signed domain with a secure delegation is a straightforward, field-proven administrative process: It can be achieved simply by renting the domain, setting up the delegation as well as an open-source authoritative DNS server, and configuring it with the zone files following the outline in this section.
To attract the victim resolver the attacker can take various approaches, e.g., by embedding image URLs in HTML documents and distributing them via e-mail or ad network, or by sending bogus e-mail to an SMTP server configured to deliver bounce messages [RFC5321].
The specific approach taken is out of scope of this document.
The resources required by the attacker are generally low.
Specifically, the attacker does not require any potent hardware, since the attacks utilize only a limited number of network transactions (the most potent one requiring only a single attack packet).
Fruthermore, the cryptographic material, which the victim resolver will be busy validating, is invalid by design, rendering it computationally trivial to generate.}

\subsection{DNSSEC Setup}
The attack vectors are encoded in a zonefile in the domain controlled by the adversary. For the attack to be effective, the adversary needs to register a domain under a signed parent. 

{\bf Zonefile.}
The attack vectors are encoded in the zonefiles in Figure \ref{fig:grid}. In addition to the DNSSEC records, the zones also feature DNS records.

{\bf Chain of trust.} Although there is no explicit requirement in the DNSSEC standard how validation of signed DNS records should proceed, the standard specification suggests that it should be done top down. The validator should construct the chain of trust top down [RFC4033, §3.1], and is required to authenticate the DNSKEY before using it to validate signatures, see [RFC4035, §5.3.1].

\subsection{SigJam (One Key x Many Signatures)}\label{sc:sigjam}
\begin{figure}[htp]
    \centering
    \begin{subfigure}{0.49\columnwidth}
        \centering
        \includegraphics[width=\columnwidth]{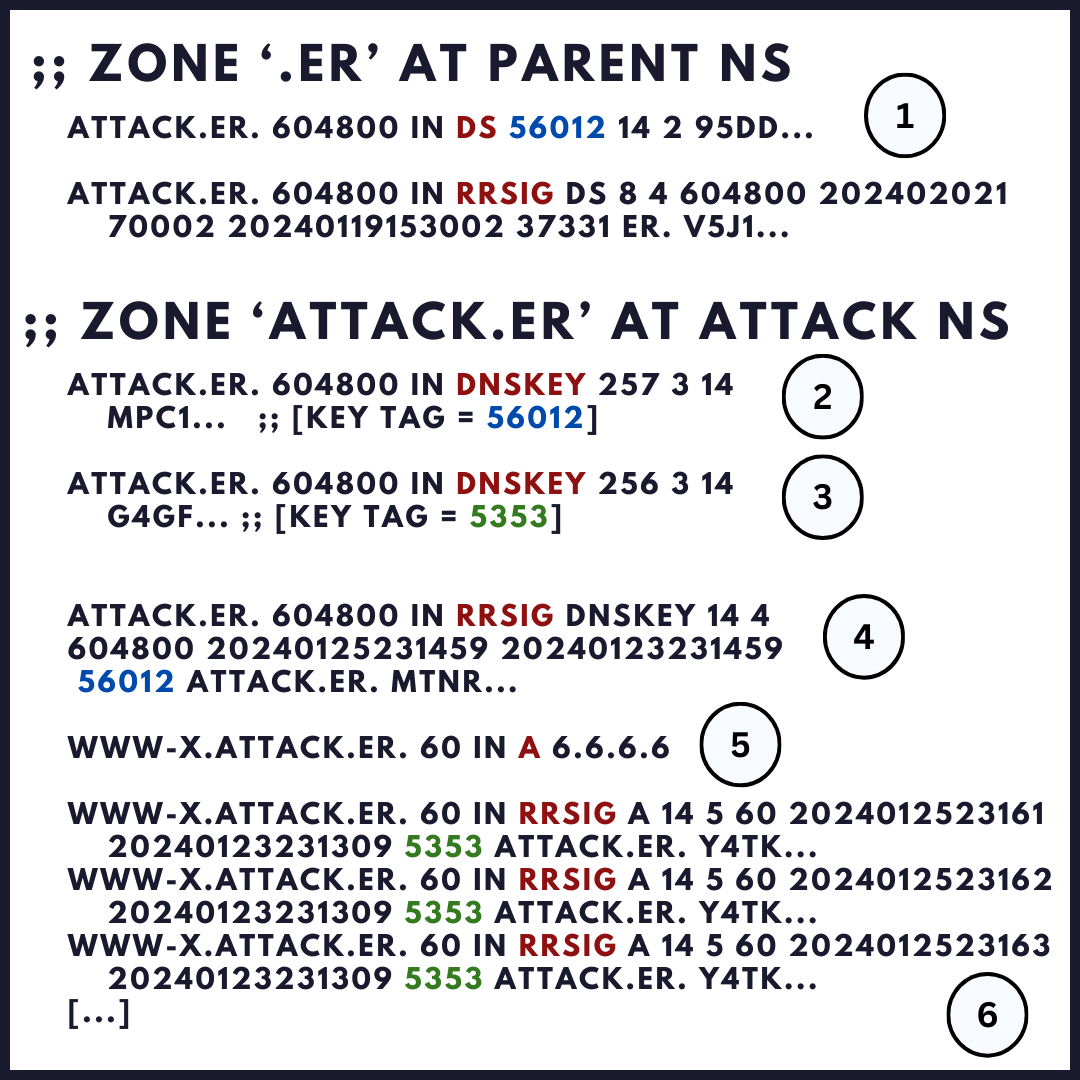}
        \caption{Zonefile for SigJam attack.}
        \label{fig:sigjam_zone}
    \end{subfigure}
    \hfill 
    \begin{subfigure}{0.49\columnwidth}
        \centering
        \includegraphics[width=\columnwidth]{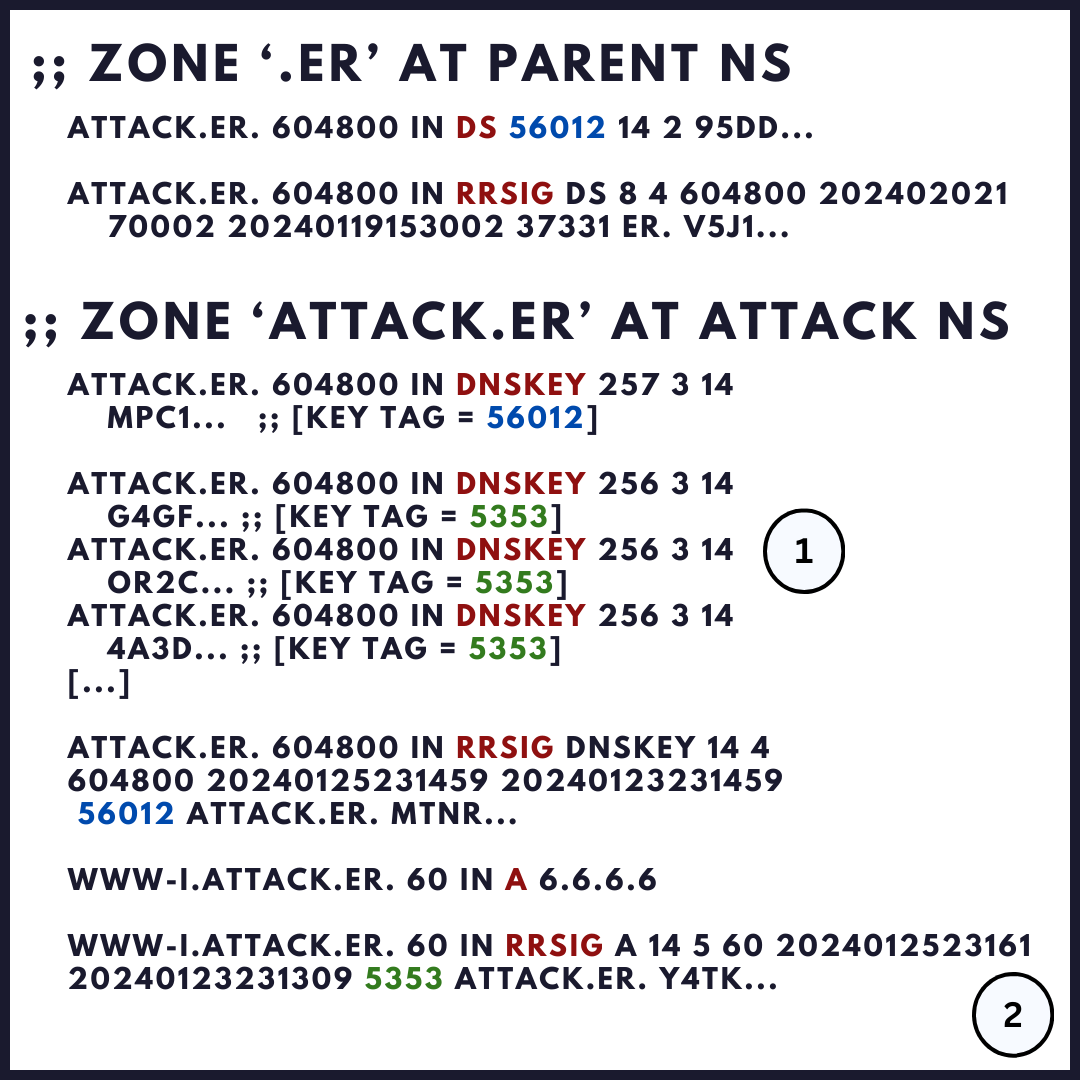}
        \caption{Zonefile for LockCram attack.}
        \label{fig:lockcram_zone}
    \end{subfigure}

    \begin{subfigure}{0.49\columnwidth}
        \centering
        \includegraphics[width=\columnwidth]{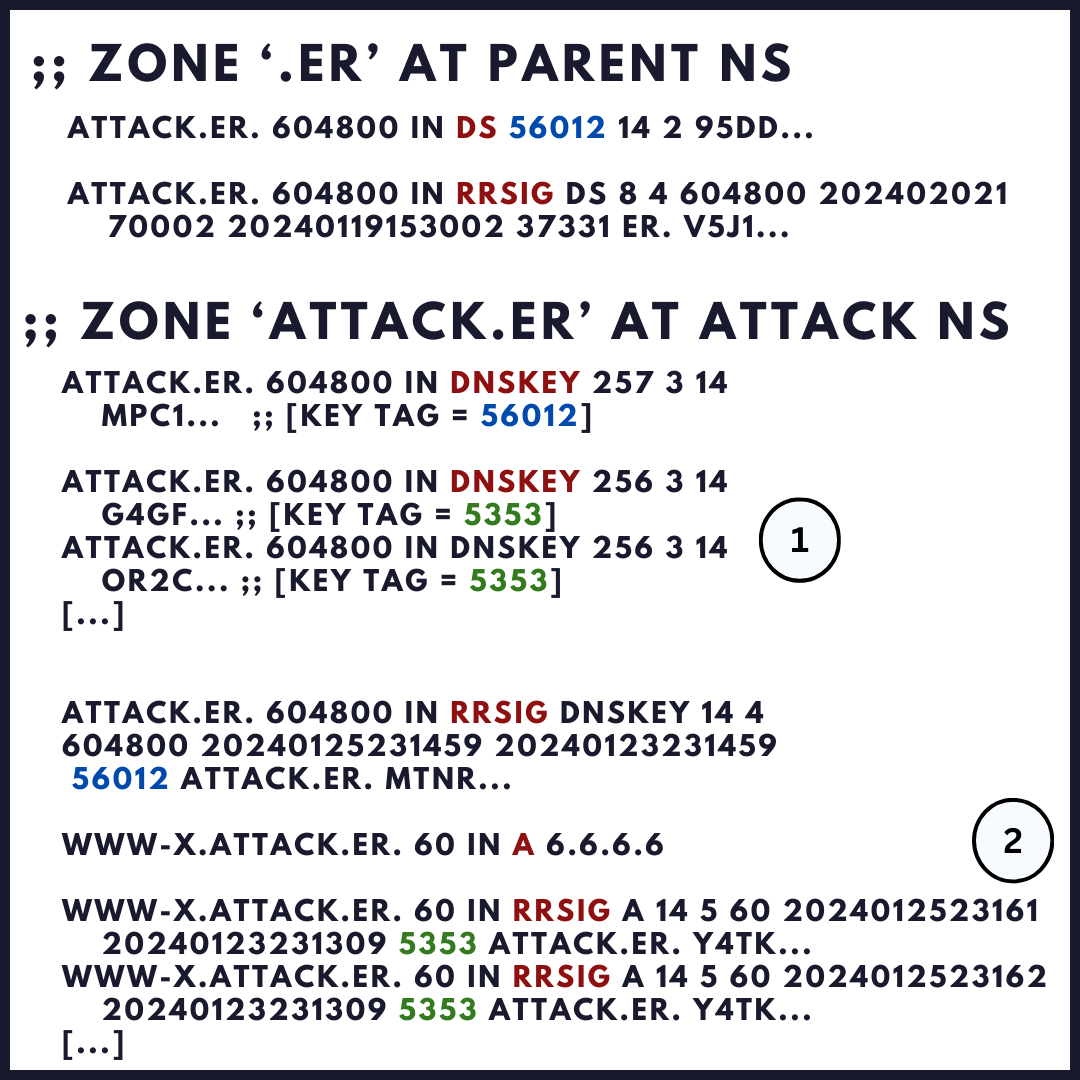}
        \caption{Zonefile for KeySigTrap attack.}
        \label{fig:keytrap_zone}
    \end{subfigure}
    \hfill
    \begin{subfigure}{0.49\columnwidth}
        \centering
        \includegraphics[width=\columnwidth]{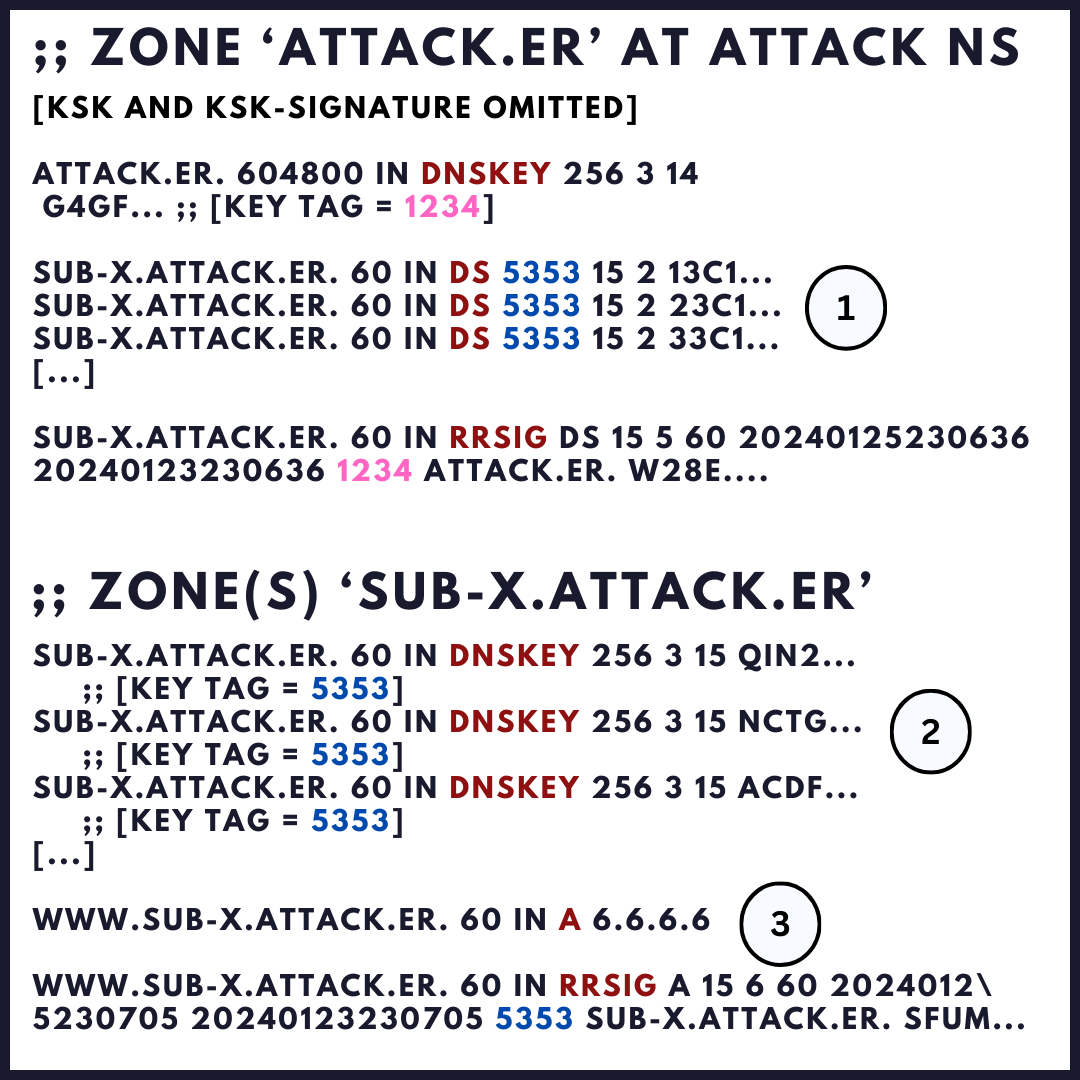}
        \caption{Zonefile for HashTrap attack.}
        \label{fig:hashtrap_zone}
    \end{subfigure}
    \caption{Zonefiles of the KeyTrap algorithmic complexity attacks.}
    \label{fig:grid}
\end{figure}

The RFC advises that a resolver should try all signatures until a signature is found that can be validated with the DNSKEY(s). To exploit this recommendation we construct an adversarial zone illustrated in Figure \ref{fig:sigjam_zone}. The parent zonefile contains a signed DS record \circled{1} that authenticates the KSK (key tag 56012) of the child \circled{2}. The zonefile of the child contains a ZSK (key tag 5353) \circled{3} signed with the KSK (key tag 56012) \circled{4}. Finally the ZSK is used to sign the A record \circled{5} with multiple (invalid) signatures all of which refer to the same ZSK DNSKEY record \circled{6}.

The RFC-compliant resolver should try all the signatures with the key until one validates or no signature is left to try. Mapping between the key and the signatures is done by matching the triple signer name (attack.er), algorithm (14), and key tag (5353). The resolver only tries the key(s) where this triple matches. However, since none of the three values needs to be unique, collisions can occur, i.e., where multiple signatures fit the same DNSSEC key. 
As indicated in the example zone by \textit{[...]}, the attacker adds many invalid signatures, all matching the triple of the ZSK. 
The resolver tries all the signatures. Since none validates, the resolver concludes that the record cannot be validated and returns a SERVFAIL error to the client, that requested the A record. The SigJam attack is thus constructed by leading the resolver to validate many invalid signatures on an A record using one ZSK.

\subsection{LockCram (Many Keys x One Signature)}\label{sc:lockcram}
Following the design of SigJam, we develop an attack vector, we dub LockCram, that exploits the fact that resolvers are mandated to try all keys [RFC4035] available for a signature until one validates or all have been tried. The LockCram attack is thus constructed by leading a resolver to validate one signature over a DNS record using many ZSK DNSSEC keys. The zonefile for the LockCram attack is illustrated in Figure \ref{fig:lockcram_zone}. Since the zone has multiple similarities to the SigJam zone, in the following we highlight the differences.

The zonefile contains multiple ZSK keys in a key resource set \circled{1}, an A record and an (invalid) signature over an A record \circled{2}. The keys can be validated by a single KSK of the zone (key tag 56012). 
The attacker needs to ensure that all the ZSKs match in signer name, algorithm and key tag. While matching signer name and algorithm is trivial, matching the key tag is not straightforward, as the key tag is not explicitly stated in a key field, but instead calculated over the record data of the key. Further, the attacker cannot simply add the exact same key multiple times, allowing identical key tags, as resolvers de-duplicate identical entries. However, the attacker can brute-force the creation of colliding keys, i.e., keys with identical key tags and differing key bits: The adversary continuously creates new DNSSEC keys with the desired algorithm, calculates the key tag and only stores keys with the desired tag until the target number of colliding keys has been collected. In the example the attacker needs to create many DNSSEC keys with algorithm 14 and the key tag 5353. All of these keys are added to the zone.

A resolver that queries the A record attempts to validate the signature. To do that the resolver identifies all the DNSSEC keys for validating the signature, which in this example are all the ZSKs, as all have a matching triple of signer name (attack.er), algorithm (14), and key tag (5353). An RFC-compliant resolver must try all the keys on the invalid signature until concluding the signature is invalid, leading to multiple validations in the resolver.

\subsection{KeySigTrap (Many Keys x Many Signatures)}\label{sc:keytrap}
The KeySigTrap attack combines the many signatures of SigJam with the many colliding DNSKEYs of LockCram, creating an attack that leads to \ccs{quadratic complexity of validations, while the other two attacks scale linearly with the number of abused records}. Figure \ref{fig:keytrap_zone} illustrates how the KeySigTrap attack zonefile can be constructed. We highlight the differences to the previous zones.

The construction of the DNSKEY set in the KeySigTrap attack is identical to the set in the LockCram attack. The attacker creates a set of ZSKs in step \circled{1}, all with the same key tag (5353). With a large amount of keys, the attacker ensures that each signature validation requires as many validations as there are colliding ZSKs in the zone.
The attacker additionally uses the idea of SigJam to put many signatures with the same key tag into the zone in step \circled{2}, which need to be validated to authenticate a DNS record, in this example an A record. This ensures that a resolver querying the A record for \textit{www-X.attack.er} must validate a large number of signatures. 

The attacker zone therefore contains many signatures over the requested hostname, and each of these signatures refers to many different ZSKs. Following the RFCs, the resolvers \should{} try all the signatures. For each signature the resolver tries, it \must{} try all the matching ZSKs in the zone. In the example zone the resolver thus tries to validate every signature with every ZSK, leading to an immense amount of signature validations, until the resolver concludes that the DNS record could not be authenticated.

\subsection{HashTrap (Many Keys x Many Hashes)}\label{sc:hashtrap}
\ccs{While mitigations may be introduced to limit signature validations, it is important to note that complexity attacks can also be achieved through hash computations.}
The concept of the attack, we dub HashTrap, is illustrated in the example zone in Figure \ref{fig:hashtrap_zone}. 
In DNSSEC, DS records from a parent zone are used to authenticate individual DNSKEY records in a child zone. This is done to delegate trust from a parent zone public key to a child zone public key. DS records use the same triple (owner name, algorithm, key tag) as RRSIGs do to identify a subset of candidate DNSKEYs. Only by validating that the hash in the DS record matches the digest of the DNSKEY can the resolver determine that a pair of DS and DNSKEY records actually belong together, which is an operation with worst-case \ccs{quadratic increase in validation complexity}, similar to the algorithm exploited in \ref{sc:keytrap}. We construct a CPU resource consumption attack, which abuses this DNSSEC protocol inefficiency.
The attacker creates additional child zones of the attacker zone, represented as \textit{sub-x.attack.er} in Figure \ref{fig:hashtrap_zone}.
For each of these sub-zones, in step \circled{1} the attacker provides numerous DS records referring to the same key tag (5353), algorithm and signer name of the DNSKEYs in the child. Our attacker utilizes unique digest values to ensure the DS records in the record set are not de-duplicated at the resolver. Since these hashes are purposefully invalid, the attacker can select arbitrary values. The record set containing all the hashes is signed with a single signature by a ZSK of the \textit{attack.er} zone.

The resolver needs to authenticate the DNSKEY records before using the keys to validate the signatures [RFC4035]. 
Before authenticating the signature over the DNSKEY set the resolver first needs to find the DNSKEY that matches a DS record from \textit{attack.er}. 
This is exploited in the attack. The attacker creates many unique DNSKEYs \circled{2}, all with the identical name, algorithm, and key tag (5353).
To find the correct key to validate with a given DS hash in the parent zone, the resolver has to iterate over all colliding keys, calculate the hash and compare it to the hash in the DS record.
This is repeated with all DS hashes in the parent zone and all DNSKEYs in the child zone, which is an operation with worst-case asymptotic complexity in $\mathcal{O}(n^2)$, leading to a substantial amount of hash calculations that mount severe computational load on the resolver. The resolver can only conclude that none of the DNSKEYs can be used to authenticate the signature \circled{3} after all the hashes were calculated. 
Our experimental evaluations show that, by means of exploiting of this attack vector, hash computation is sufficiently resource intensive to inflict excessive load on a resolver.

The HashTrap attack is thus constructed by leading the resolver to calculate many hashes for validating many colliding DNSKEYs against many DS hash records. Notice that attack variants, similar to SigJam and LockCram, can also be constructed with the DS hashes instead of signatures. However, hashes are less effective, reducing adversary's motivation to do that.

\section{Evaluation of the Attacks}\label{sec:praceval}

Through experimental evaluations we found all the major DNS implementations on our dataset to be vulnerable to KeyTrap attacks. The stalling interval caused by the attacks depends on the specific resolver implementation. Our list of DNS software includes recursive DNS resolvers, public resolvers\footnote{Tested against an instance set up by the operator for this purpose}, DNS tooling, and DNS libraries; see details in Table \ref{tab:vulnerable:implementations:new}.
We consider a resolver vulnerable if we achieve full DoS with traffic <10 req/s.
\ccs{All evaluations were conducted on an Intel Core i7-8650U quad-core processor with up to 4.2 GHz single core frequency.}
We describe the setup, our test methodology, and the cryptographic ciphers we use in our research zonefiles. We then evaluate the effectiveness and the impact of the attacks.
An overview over the different DNS resolver components relevant to the attack is given in Table \ref{tab:components}.
\ccs{Note that Knot, unwind, and dnsmasq have tight response buffer size limitations, unintentionally reducing the impact of the attack. In fact, this side effect was recognized as a bug in Knot and fixed in the patched version of the resolver \footnote{https://github.com/CZ-NIC/knot-resolver/commit/\\0b8012c2d68b7d59a55a0dca1d3f0c3042016ae9}.}

\begin{table}[t!] 
\centering 
\begin{center}
\renewcommand{\arraystretch}{0.7}
   \scriptsize
\begin{tabular}{|r|l|c|p{3cm}|}
    \hline
    \ignore{\rot{Type}} & \textbf{Name}               & \textbf{Vuln.} & \textbf{Comment}   \\
    \hline
    \multirow{17}{*}{\rot{\textbf{Server Software}}}
        
         & Akamai CacheServe & \cmark  & DoS 186s  \\
         & BIND9                 & \cmark & DoS 58632s\\
         & Knot Resolver & \cmark & DoS 51s (Limited DNS key buffer size)\\
         & PowerDNS Recursor & \cmark & DoS 177s \\
         & Unbound & \cmark &  DoS 1014s (Retries increase duration) \\
         & Windows Server 2022 & \cmark & DoS 131s \\
         & Windows Server 2019 & \cmark & DoS 134s \\
         & unwind (from OpenBSD7.3) & \cmark & DoS 29s (Limited msg-buffer size) \\
         & Technitium &   \cmark  & DoS 411s \\
         & dnsmasq 2.80                & \cmark & DoS 2s (Limited msg-buffer size) \\  
         & stubby 0.4.3           &    \cmark & DoS 184s \\
    \hline
    \multirow{4}{*}{\rot{\textbf{Service}}}  
         & Cloudflare              & \cmark &  Confirmed by Developers \\
         & Google              & \cmark & Confirmed by Developers  \\
         & OpenDNS              & \cmark & Confirmed by Developers    \\
         & Quad9              & \cmark &   Confirmed by Developers \\
    \hline
    \multirow{7}{*}{\rot{\textbf{Tool}}}
         & dig 9.16.1               &    \xmark &  No DNSSEC validation               \\
         & kdig 2.7.8             &    \xmark & No DNSSEC validation                \\
         & delv 9.16.1              &    \cmark & Validation logic from Bind9                \\
         & DNSViz 0.9.4 (latest)  &    \cmark & Crashes with exception after attack                \\
         & ldns-verify-zone & \cmark & uses vulnerable ldns library \\
         & kzonechek & \cmark & shipped with Knot DNS authoritative server \\
         & named-checkzone & \xmark & does not validate signatures \\
         
    \hline
    \multirow{4}{*}{\rot{\textbf{Libs}}}
         & dnspython              &    \cmark & - \\
         & getdns                & \cmark & used by stubby               \\
         & ldns                & \cmark & -                   \\
         & libunbound              &    \cmark &  used by Unbound        \\
   
    \hline
\end{tabular}
\end{center}
\caption{Vulnerable DNS implementations.}
\label{tab:vulnerable:implementations:new}
\end{table}

\subsection{Setup}
Unless mentioned otherwise, all evaluations are run on a single CPU core. This allows us to compare between different resolvers with various multi-threading standard configurations. We set up a test environment with a number of components.

{\bf Components.} 
We set up resolvers and DNS servers in an isolated environment. This ensures that attack requests are not propagated to validating upstream DNS resolvers. 
We develop scripts for automated dynamic generation of zonefiles and records upon each query, and scripts for automated construction of the DNSSEC chain of trust. 
Generating the zonefiles dynamically enables us to use a virtually infinite number of zones and records required for testing the attacks, which would have otherwise quickly cluttered the zone files and hampered investigations.
The nameservers host the domains used for testing the resolvers and exchanges DNS messages with them according to protocol specifications and specific test semantics. Each test is hosted in a separate subdomain consisting of one or multiple zones. This prevents cache-induced interference between consecutive executions of tests and reduces implementation complexity of the investigations. Test configurations are pre-generated from configuration templates, which we define using a small domain-specific language.
This allows efficient variations over the signature algorithms or the specific number of RRSIGs and DNSKEYs in responses, which are provisioned for attacking validation routines.
We conduct tests by sending queries to the resolvers, causing them to interact with our nameservers according to the test configurations. When a nameserver receives a query it parses it, matches it against a pre-defined set of rules and generates a response. The rule set is loaded from a configuration file upon startup, and determines which tests can be conducted, as well as the specific test semantics. A "test" specifies, e.g., a set of domains with specific DNSSEC algorithms, numbers of DNSKEYs and signatures over records to validate against these DNSKEYs.

{\bf Transport protocol.} DNS responses are typically delivered over UDP. When DNS responses are too large, e.g., exceeding the EDNS size in EDNS(0) OPT header, the nameservers fall back to TCP to avoid fragmentation. 
Our attacks can be implemented either over UDP or TCP. We implement TCP as the transport protocol between the resolvers and our nameservers. The maximum size of a DNS message sent over TCP is dictated by [RFC1035], stating that a DNS message over TCP must have a length value prefixed to the message with 2-octets size. Resulting from the size limitation of this field, DNS payload sent in a response from the nameserver to the resolver can have a maximum
size of $2^{16}-1$ = 65535 bytes.
Depending on the Maximal Transmission Unit (MTU), this payload will be sent in one or more TCP segments. 
Therefore, the attack payload (i.e., DNS/DNSSEC records) in a DNS response is limited to 65K bytes.

\subsection{Identifying the Optimal Cipher for Attacks}\label{sec:derviation_amount}
Different DNSSEC algorithms vary in the mathematical computation logic and the complexity of mathematical operations. Therefore, the computation of DNSSEC validation for different algorithms differs in the amount of computation time. 
This means that the load created by our attacks is determined also by the cryptographic ciphers the adversarial domain uses. DNSSEC generally supports two different algorithm suites\footnote{\path{https://www.iana.org/assignments/dns-sec-alg-numbers/dns-sec-alg-numbers.xhtml}}: RSA based and Elliptic Curve Cryptography (ECC) based cryptographic algorithms. We evaluate both suites and find that the ECC based cryptographic algorithms exhibit a significantly higher load than RSA based algorithms and surpass RSA by over an order of magnitude, \ccs{even when considering RSA keys with the most inefficient selection of exponents allowed by the DNSSEC specification.}
This is consistent with previous work \cite{van2016performance}. ECC-based algorithms are thus better suited to maximize the impact of our attacks on resolvers. We therefore focus on ECC-based algorithms in the following evaluations of the attacks.

\begin{table}[t!]
\renewcommand{\arraystretch}{0.7}
   \scriptsize
\begin{tabular}{l|rrr}
\toprule
 & Keys & Signatures & Validations \\
Cipher &  &  &  \\
\midrule
ED448 & 907 & 454 & 411\,778 \\
ED25519 & 1\,391 & 696 & 968\,136 \\
ECDSAP384SHA384 & 589 & 519 & 305\,691 \\
ECDSAP256SHA256 & 828 & 696 & 576\,288 \\
RSA-512 & 788 & 696 & 548\,448 \\
RSA-1024 & 444 & 413 & 183\,372 \\
RSA-2048 & 237 & 228 & 54\,036 \\
RSA-4096 & 122 & 119 & 14\,518 \\
\bottomrule
\end{tabular}
\caption{Max. DNSKEYs, RRSIGs, and validations per response.}

\label{tab:maximum-dnssec-rr-count}
\end{table}

{\bf Comparison of computation load of ECC algorithms.} 
We evaluate if validation of different ECC algorithms results in different processing times on different DNS resolvers. For the evaluation, we set up all major DNS resolvers (see Table \ref{tab:ecc}) on an identical hardware machine. We evaluated all resolvers by running a full resolution with 2500 validations. Times were average over 10 attempts to ensure consistency. Measuring the validation time of the message instead of only measuring the validation procedure allows a more accurate view of the behavior of the resolvers for different algorithms, as overall processing times might also be influenced by components outside the validation procedure. The measurements illustrate different processing times between resolvers, indicating differing efficiencies of the implementation. Some efficiency divergence is expected, as a large amount of signature validations on a single RRset is not an expected use-case for resolvers and thus, it is expected that resolvers are not optimized for it. This is clearly visible in the validation times of Bind9, which supersede the other resolvers due to an inefficient implementation of key selection in the case of colliding keys.

The table illustrates that all resolvers take the longest validation time for signature created with algorithm 14, \ccs{which is} ECDSA Curve P-384/SHA-384. 
Thus, algorithm 14 is the most suited for the attacks on all resolvers, achieving maximum impact with the available maximum buffer size. Using the 384bit key size of algorithm 14, and constructing the theoretical minimal size DNS message transporting the keys, an attacker could fit up to \textbf{589 colliding DNS keys into a single DNS message}. Similarly, using minimal DNS overhead, an attacker could fit up to a maximum of \textbf{519 signatures into a single DNS message}. Thus, with one resolution request with algorithm 14, an attacker could theoretically trigger \textbf{589*519 = 305691 signature validations} in the DNS resolver, leading to significant processing effort on the resolver. Table \ref{tab:maximum-dnssec-rr-count} shows the theoretical maximum values for all commonly supported DNSSEC algorithm. Theoretical maxima are calculated by choosing the minimal possible size for all fields in the DNS message. In practice, the exact value of signatures and keys that can fit into a single message is limited by the attacker setup. DNSSEC messages contain additional information besides the raw bytes of the signature or key, like the signer name, leading to the lower number of entries in real-world attack setup. In our evaluation setup, we make a conservative approximation on the practical size of the fields in the messages, using 582 DNSKEYs per message and 340 signatures per message.

\begin{table}[t!]
\renewcommand{\arraystretch}{0.7}
\centering 
\scriptsize
\begin{tabular}{|l|c|c|c|c|}
\hline
Name & Alg 13 & Alg 14 & Alg 15 & Alg 16 \\ \hline
Unbound & 172 & \textbf{996} & 880 & 364 \\ 
Bind9 & 888 & \textbf{2448} & 460 & 628 \\ 
Knot & 232 & \textbf{496} & 164 & 456 \\ 
Akamai & 219 & \textbf{976} & 209 & 392 \\ 
PowerDNS & 153 & \textbf{924} & 840 & 628 \\ \hline
\end{tabular}
\caption{Validation time per signature in $\mu$s.}
\label{tab:ecc}
\end{table}

\subsection{Effectiveness of the Attacks}
To evaluate the attacks, we setup all the resolvers to query a malicious domain signed with algorithm 14.
During the evaluations, we use a benign DNS client that requests ten unique DNS entries per second from the investigated resolver and logs received replies. We choose a 5s timeout for benign requests, i.e., benign requests to the resolver that are not answered after 5s are considered to have no value to the benign user and are therefore considered lost. This timeout is consistent with DNS tooling like dig (5s) \footnote{https://linux.die.net/man/1/dig}, Windows DNS tools (1s-4s), and glibc (5s) \footnote{https://linux.die.net/man/5/resolv.conf}.

\begin{figure}[b!]
    \centering
    \includegraphics[width=0.9\columnwidth]{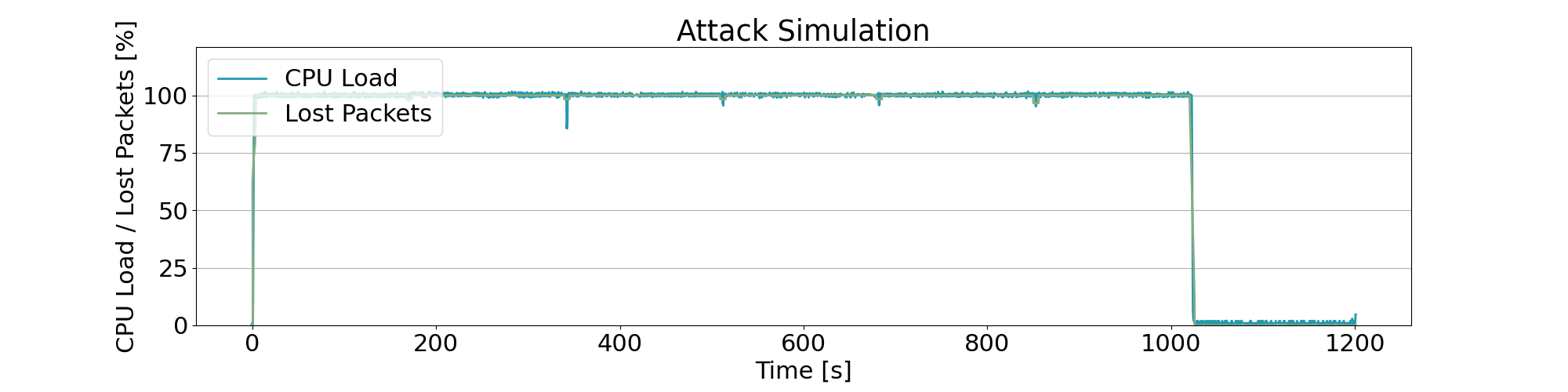}
    \caption{KeySigTrap attack on Unbound with single request.}
    \label{fig:unbound_hpoc}
\end{figure}

{\bf KeySigTrap.} Evaluating KeySigTrap, we set up a zonefile with 582 colliding DNSSEC keys and 340 signatures. 
We illustrate the impact of the attack on Unbound in Figure \ref{fig:unbound_hpoc}. As can be seen in the plot, once the attacker triggers a single DNS request, the KeySigTrap attack payload in the DNS response causes the CPU usage on the resolver to increase to 100\% due to a large load in validating the signatures. While busy validating signatures, the resolver does not answer any benign requests, leading to 100\% lost benign requests until the resolver finishes the validation, which takes about 1014s. Thus, a single attacker request causes a 1014 seconds long complete DoS of the resolver. We measured all investigated DNS resolvers on an identical setup. The results in Table \ref{tab:hpocsq} show that all resolvers are heavily affected by a single request and stalled for a substantial amount of time. However, the stalling duration differs significantly between resolvers. Akamai, PowerDNS and Stubby all take about 3 minutes to validate the signatures. The reason is that they use similar cryptographic implementations, validating through all key-signature pairs until they return a SERVFAIL to the client. \ccs{We find that on average, a KeySigTrap request causes 2.000.000x load compared to benign requests, reducing resolver throughput to 0 in any tested resolver.}
However, we observed three notable outliers in the DoS duration of the attack.

\textit{Unbound} is DoSed approximately six times longer than the other resolvers. The reason is the default re-query behavior of Unbound. In a default configuration, Unbound attempts to re-query the nameserver five times after failed validation of all signatures. Therefore, Unbound validates all attacker signatures six times before returning SERVFAIL to the client. 
This explains the extended DoS of Unbound compared to the other resolvers. Disabling default re-queries, we find Unbound is DoSed for 176s on a single KeyTrap request.  

\textit{Bind9} is the second major outlier. The resolver is stalled for over 16h with a single attacker request. Investigating the cause for this observation, we identified an inefficiency in the code, triggered by a large amount of colliding DNSSEC keys. The routine responsible for identifying the next DNSSEC key to try on a signature does not implement an efficient algorithm to select the next key from the remaining keys. Instead, it parses all keys again until it finds a key that has not been tried yet. The algorithm does not lead to inefficiencies in normal operation with a small amount of colliding keys. But when many keys collide, the resolver spends a large amount of time parsing the keys and selecting the next key, extending the duration of the DoS to 16h.

\textit{Knot} is slightly less affected by the attack than the other resolvers. Evaluating the attack on Knot shows that the resolver has a limited buffer size for DNSSEC keys, limiting the number of keys per request to 126 keys. This results in a shorter DoS duration on Knot. However, the impact of the attack on Knot is still substantial with a 56s DoS from a single attack request.

In the following, we will show the impact of SigJam, LockCram, and HashTrap on the resolvers, illustrating how to similarly achieve maximum DoS of the resolver.

\begin{figure}[t!]
     \centering
     \includegraphics[width=0.9\columnwidth]{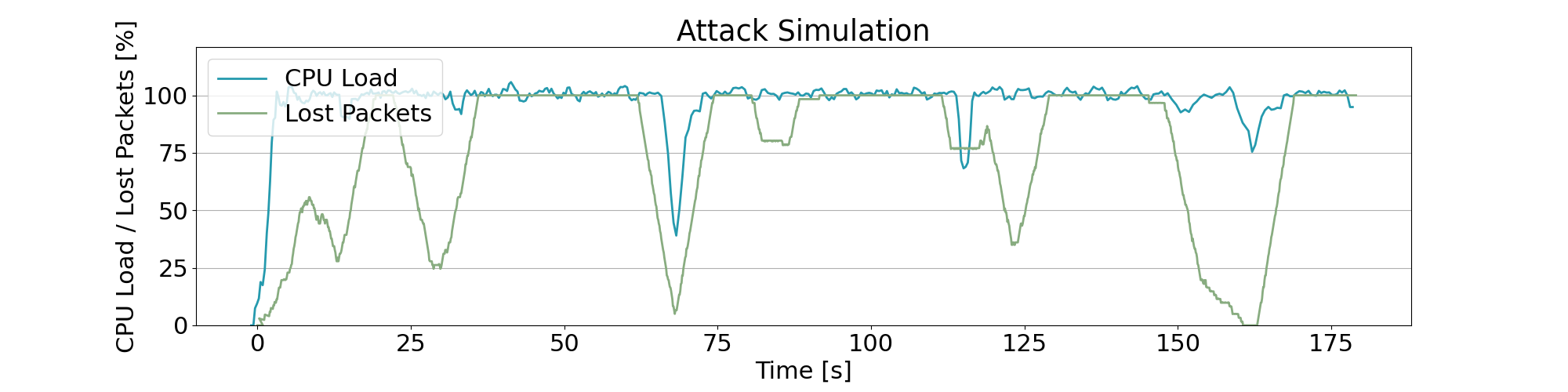}
     \caption{SigJam attack with 10 req/s.} 
     \label{fig:sigjam}
 \end{figure}

{\bf SigJam.} 
Achieving full DoS with any attack other than KeySigTrap requires more than a single attacker request. To evaluate SigJam, we send 1 attack response per second to the resolver, containing the (maximum number of) 340 signatures in one DNS response. 
Using 340 signatures per request, we steadily increase the amount of attacker's requests until we observe no increase in lost benign queries. As illustrated in Figure \ref{fig:sigjam}, 10 req/s cause a severe load on the resolver, leading to 75\% lost benign traffic. The reason for intermediate responses to benign queries is I/O when the resolvers wait for new signatures. \ccs{We find that on average, a SigJam request is able to displace 773 benign requests, i.e. a SigJam request causes 773x load compared to begin traffic. For a resolver capable of handling 1000 requests in real-time, the throughput is reduced to 227 req/s with a single SigJam request per second.}  
This also explains why we do not see improvement in effectiveness of the attack with more malicious requests. The resolver still needs to conduct I/O operations, hence intermediate requests get processed. 
 
 \begin{figure}[b!]
    \centering
    \includegraphics[width=0.9\columnwidth]{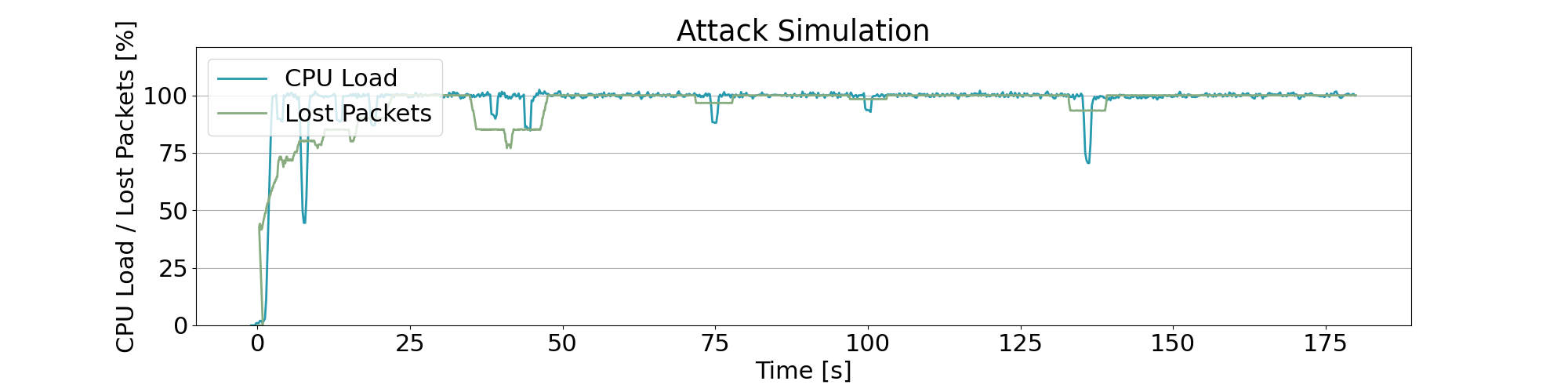}
    \caption{HashTrap attack with two requests per second.}
    \label{fig:hashtrap}
\end{figure}  

\begin{figure}[t!]
     \centering
     \includegraphics[width=0.9\columnwidth]{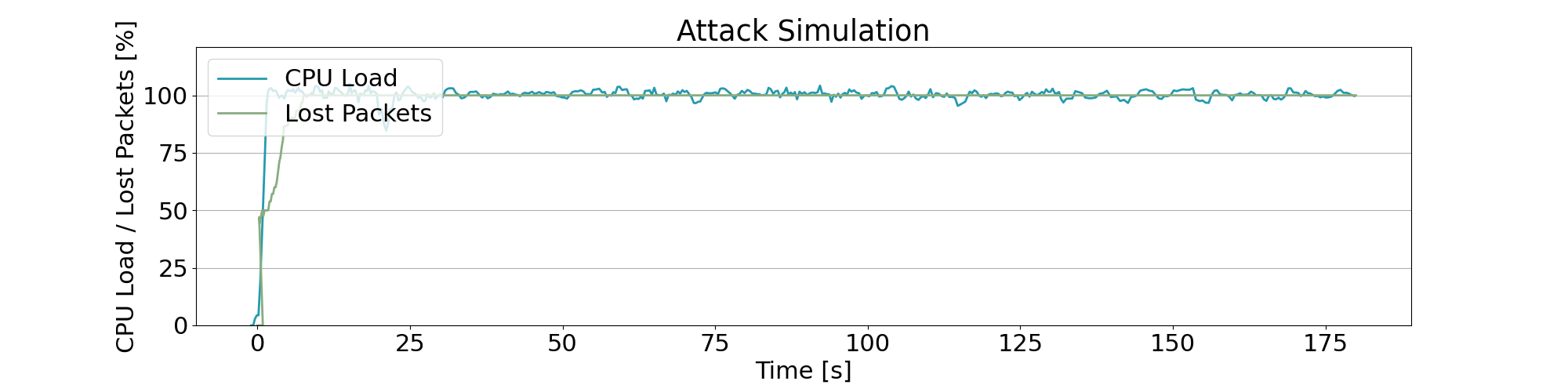}
     \caption{LockCram attack with 10 req/s.}
     \label{fig:lockcram}
 \end{figure}

{\bf LockCram.} We evaluate the LockCram attack using 582 keys of algorithm 14 on Unbound. 
The attack starts with 1 attacker request per second. We increase the rate of attack until we see no increase in lost benign requests. 
At 10 attack req/s, we achieve full DoS of the resolver, with > 99\% loss of benign requests, see Figure \ref{fig:lockcram}. 
The figure illustrates that the validation of the signature against all colliding keys results in 100\% utilization of the CPU. 
In contrast to SigJam, we do not see intermediate replies while the attack is running. The reason is that LockCram attack requires much lower I/O effort than SigJam. In the first attack request of the evaluation, the resolver needs to download and validate the RRSet containing all colliding keys. In subsequent requests, the resolver already has the keys cached and only needs to download one signature. Thus, the resolver spends much less time idling during the attack, preventing it from answering benign requests while waiting for attack I/O. \ccs{On average, a LockCram request is able to displace 815 benign requests, i.e. a LockCram request causes 815x load compared to benign requests.}

{\bf HashTrap.} We evaluate the HashTrap attack using digests of type 2 (SHA256) as it requires the largest amount of time to compute on a 64-bit system. 
Since the calculation time of the hashes does not depend on the key size, we chose the smallest possible DNSSEC keys, fitting as many keys as possible in one DNS message and thereby maximizing the number of hash calculations. The smallest key size of common DNSSEC algorithms is given by algorithm 15, using 256-bit keys. 
Using 256-bit keys allows us to fit 1357 DS records, and 1357 DNSKEYs in one attack request, resulting in 1357 * 1357 = 1.841.449 hash calculations per request.

We start the evaluation with 1 attack request per second and increase the rate until we observe no further increase in lost benign requests at 2 attack requests per second. As can be seen in Figure \ref{fig:hashtrap} the attack leads to 98\% lost benign request. The 2\% queries still answered are again caused by I/O operations of the resolver, allowing it to answer some benign queries. \ccs{We find a greater impact on maximum throughput, with one HashTrap request displacing 1254 benign requests.}

\begin{table}[b!]
\renewcommand{\arraystretch}{0.7}
\centering 
\scriptsize
\begin{tabular}{|l|c|}
\hline
Name & DoS Duration\\ \hline
Unbound & 1014s \\ 
Bind9 & 58632s \\ 
Knot & 51s* \\ 
Akamai & 186s \\ 
PowerDNS & 170s \\ 
Stubby & 184s \\ \hline
\end{tabular}
\caption{DoS duration with single attack request.}
*Knot has a limited buffer for DNSSEC keys (126 keys), allowing for a smaller attack payload.
\label{tab:hpocsq}
\end{table}

\subsection{Effect on Inbound/Pending DNS Packets}\label{sc:udp:buffer} 
When resolvers are stalled from our attacks, they can neither process pending requests nor respond to clients queries, even for records that could otherwise be responded from the cache. 
We find that a query that arrives during the time that a resolver is stalled is generally not discarded but is placed in a buffer for later processing. In normal operation, the resolver continuously reads requests from the buffer and processes them, either by replying from cache or with a recursive DNS resolution. During a KeyTrap attack, resolvers are stalled in validation and do not process new requests. The requests are stored in the OS buffer, which eventually fills, resulting in loss of subsequent inbound packets.

Note that packets may also get lost even if the buffer is not full. We find that PowerDNS discards old packets by default. When depleting the OS UDP buffer after the attack is over, PowerDNS discards any packets older than 2s. 
This means that during the KeyTrap attack, any benign request arriving at PowerDNS earlier than 2s before the end of the attack does not get answered. If the OS buffer fills up more than 2s before the attack is over, the OS drops the packets that PowerDNS would still answer to, resulting in PowerDNS not sending out any replies to benign requests after the attack is over.

\subsection{Effect on Clients} 
We also monitor the responses sent by the resolver to a benign DNS client during the attack. The client continuously requests unique un-cached records from the tested resolver and logs when it receives an answer. With this setup, we can evaluate if the resolver still answers to benign requests while busy validating the signatures from the attack request. 

\begin{figure}[b!]
    \centering
    \includegraphics[width=0.9\columnwidth]{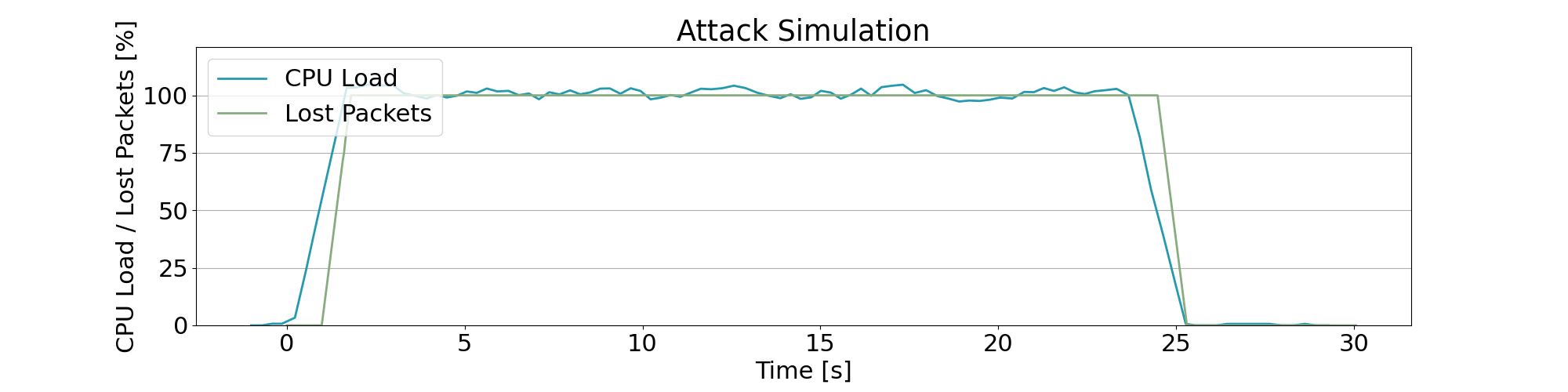}
    \caption{Impact on benign requests in Unbound under attack.}
    \label{fig:recpack}
\end{figure}

The impact is illustrated in Figure \ref{fig:recpack}. In Unbound as well as in all other resolvers we investigated, the resolver does not answer to client requests while busy validating the signatures of the attacker request. This can be seen in the graph, showing the amount of answers the client receives over time. Once the attack request is sent at two seconds, the resolver stops answering to any benign requests. Only after it finishes processing the attacker request, the resolver again answers to benign queries at around 25s. The graph illustrates that the impact of the attack is severe, as it results in a full DoS of the resolver while the attack is running.

\begin{table}[t!] 
\renewcommand{\arraystretch}{0.7}
\centering 
\scriptsize
\begin{tabular}{|c|c|c|c|c|c|c|}
\hline
Resolver  & OS Buffer & Discard & Reply & Retries & Processing  & Multithreading\\
  & Fills & old packets & to cached & &  Order & \\ 
\hline
Unbound & Y & N & N & 5 & Mesh & Internal (L/I)\\ 
Bind9    & Y & N & N & 0 & Mesh & Internal (L/I)\\ 
PowerDNS & Y & Y & N & 0 & Mesh & Internal (L/I) \\ 
Knot     & Y & N & N & 0 & Seq. & OS (L/I)\\ 
Stubby & Y & N &  N & 0 & Mesh & OS (L/I) \\ 
Akamai & N & N & Y & 0 & Mesh & Internal (L/D) \\ \hline
\end{tabular}
\caption{Components in resolvers.}
L/I: Load-Independent, L/D: Load-Dependent
\label{tab:components}
\end{table}

\subsection{Multi-Threading} 
\begin{figure}[t!]
    \centering
    \includegraphics[width=0.9\columnwidth]{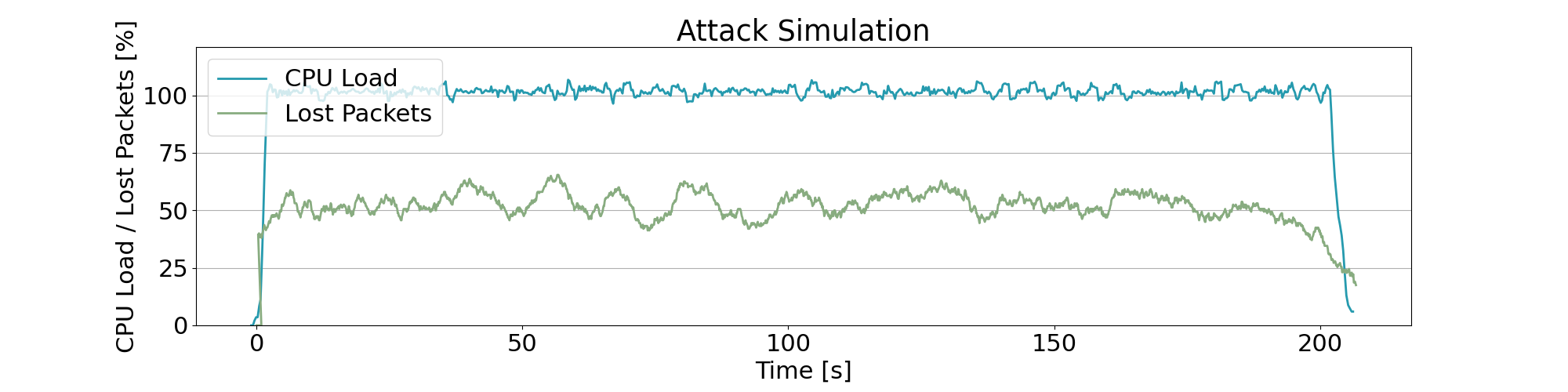}
    \caption{KeyTrap attack on Unbound with 1 request and 2 threads.}
    \label{fig:mt_unbound}
\end{figure}

Multi-threading is supported by all major DNS resolvers and influences how KeyTrap attacks affects their response behavior. To investigate the influence of multi-threading, we set up all resolvers with multi-threading enabled. Figure \ref{fig:mt_unbound} illustrates the influence of multi-threading on the attack. When using additional threads, the resolver is still able to answer to some benign requests, even while busy validating the signatures. Code review shows that the resolvers do not consider the load on a thread for scheduling, which explains why approximately half of the requests are still scheduled on a thread that is busy validating signatures. These requests are lost. Answering benign requests while validating signatures extends the duration that the resolver takes to complete validation by a short amount, in the case of Unbound by about 20s. Note that due to inherent pseudo-randomness in the scheduling of requests to the threads, and the scheduling of different threads to run by the OS, a small fluctuation of the percentage of lost requests can be observed in the graph. We observe similar fluctuations in all resolvers. We find one resolver, Cacheserve by Akamai, that does not lose parts of its traffic when multi-threading is deployed. The reason is that it considers thread load in the allocation of new requests to worker threads, leading to no lost benign requests while Cacheserve has open threads not busy validating attack signatures.

\begin{figure}[t!]
    \centering
    \includegraphics[width=0.9\columnwidth]{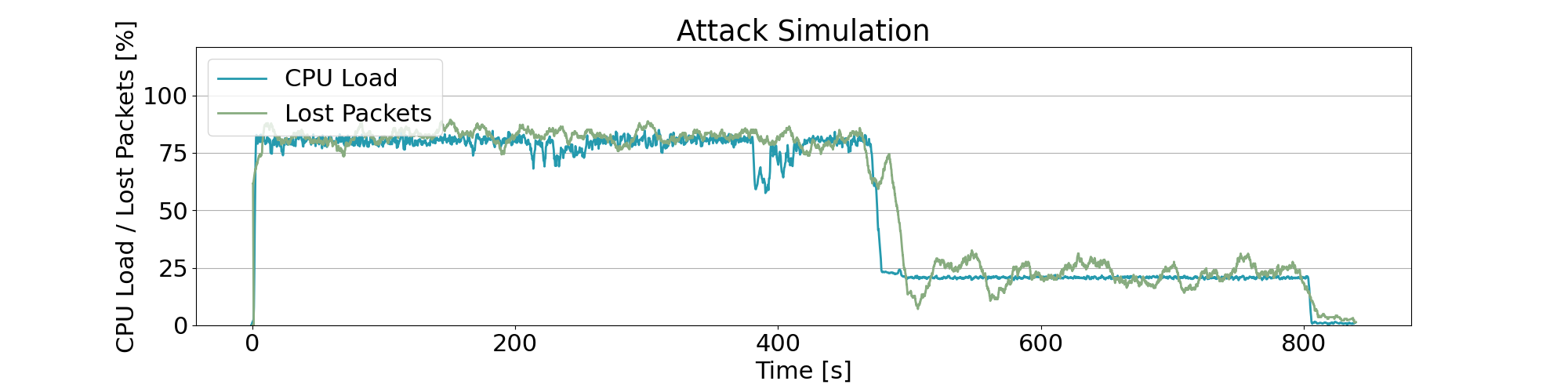}
    \caption{KeyTrap attack on Unbound with 5 requests and 5 threads.}
       \vspace{-10pt}

    \label{fig:mt5_unbound}
\end{figure}

The attacker can circumvent the supposed protection from multi-threading by sending multiple requests to the resolver. In the case of Akamai, the scheduling algorithm that considers the load of threads still allows the attacker to fill all threads with the attack. Since every new attacker request will be scheduled to a free thread, the attacker only needs to send as many attacker requests as there are threads in Cacheserve. No request will be scheduled to an already busy thread. In contrast, for all other resolvers, the success of the attack is influenced by the pseudo-random scheduling algorithm. Since allocation of requests to threads is not known to the attacker, the attacker needs to send more requests than there are threads in victim resolver to ensure all threads are hit, even if the scheduling algorithm, by chance, schedules multiple attack requests to the same thread. In the case of fully random scheduling, the average amount of attack requests needed to fill all victim threads can be calculated by
$E = n \times \sum_{i=1}^{n} \frac{1}{i}$ where n is the number of threads in the resolver. Since schedulers are usually optimized to distribute systematically to the threads, the real world average number of requests required to hit all threads will generally be lower than the random value. The effect of sending multiple queries can be seen in Figure \ref{fig:mt5_unbound}. The graph shows a scenario where the attacker sends five attacker requests to an instance of Unbound running with five worker threads on five CPU cores. As seen in the graph, the five requests do not suffice to saturate the threads, as one threads remains active in replying to benign queries, leading to approximately 80\% lost requests. The fact that two attacker requests were scheduled to the same thread can be observed in the second half of the plot. While the validation finishes in three threads, reducing the rate of lost requests by 60\%, one thread continues validating signatures for almost twice as long, indicating that two requests were scheduled to a single thread.

These observations show that multi-threading is no sufficient protection against the attack, as the attacker, when sending a sufficient amount of attack requests, can hit all threads of the resolver, leading to a comprehensive DoS of the resolver. It also illustrates that one attack request is not sufficient for a complete DoS of the resolver when multi-threading is used.

\subsection{Cached Entries} 
DNS resolvers implement a cache to answer recently requested entries without recursive resolution. This greatly improves efficiency of the resolver, as certain domains are requested more frequently, like domains of commonly used websites. However, since all the resolvers, except CacheServe, handle replies to cached entries on the same thread as recursive resolution and validation, caching does not mitigate the attack. 
In contrast, since CacheServe implements a separate thread for answering cached entries, the effect of the attack is partially mitigated.

\subsection{Continuous KeySigTrap Attack}
Using the insights gained from the previous sections, we construct a continuous attack on resolvers. 
In the initial phase of the attack, the attacker sends multiple KeySigTrap requests simultaneously. Sending multiple requests ensures that the resolver gets stalled for a substantial amount of time and, in the case of multi-threading, all threads get hit with an attack and are busy validating signatures. The DNS implementations we tested in this work use 2-6 resolution threads, depending on the resolver and the size of the deployment. Creating a real-world scenario, we thus evaluate our continuous attack on an Unbound instance running with 4 resolution threads.

The requests should be timed in such a way that new requests are always already in the buffer once a request from the previous batch finishes. Using the validation time of a single attack request in Unbound, not considering re-tries, we find a single request approximately stalls a thread for about 176s (see Table \ref{tab:hpocsq}). We choose an interval half of this duration. We further send 12, three requests per thread of the resolver, to ensure all validation threads are hit with the attack. 
The attack uses the following steps:

{\scriptsize{
\begin{verbatim}
1. Send a batch of 12 initial attack requests
2. Wait 1s to ensure the batch has been read
3. Send a batch of 12 follow-up attack request + a buffer filler
4. Wait 90s
5. Go back to 3.
\end{verbatim}
}}

\begin{figure}[b!]
    \centering
    \includegraphics[width=0.7\columnwidth]{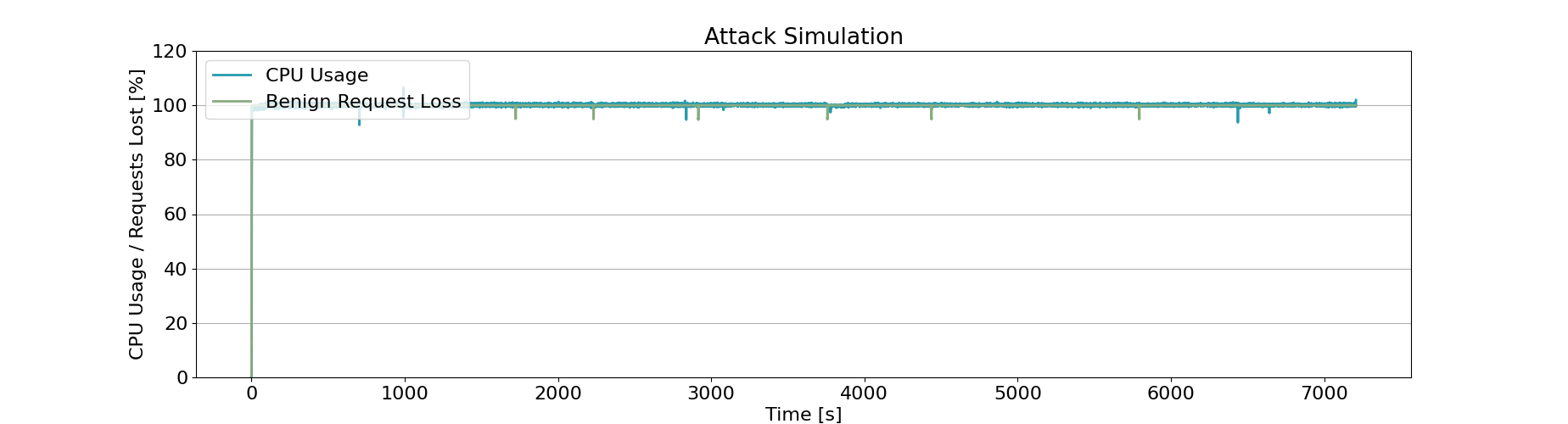}
    \caption{Continuous KeyTrap attack on 4-threaded Unbound.}
    \label{fig:2hunbound}
\end{figure}

The result of this attack is plotted in Figure \ref{fig:2hunbound}. The attack achieves a complete DoS of the resolver for the entire 2h measurement duration, with 99.999\% of benign requests lost. All 4 processor cores continuously run on 100\% CPU utilization, validating the signatures. The attacker only requires traffic of 13 request per 90s, i.e., on average one request every 6.9s. This attack rate is low enough to prevent any rate-limiting mechanisms from blocking follow up attacker requests in a real-world setting.

This evaluation demonstrates that KeySigTrap is a practical attack, achieving a continuous DoS even on a multi-threaded resolver. Even a small-scale attacker can exploit KeySigTrap to fully stall DNS resolution in the resolver for other clients for an indefinite amount of time. 

\section{The Path to Mitigations}\label{sec:countermeasures}
\begin{figure}[t!]
    \centering
    \includegraphics[width=1\columnwidth]{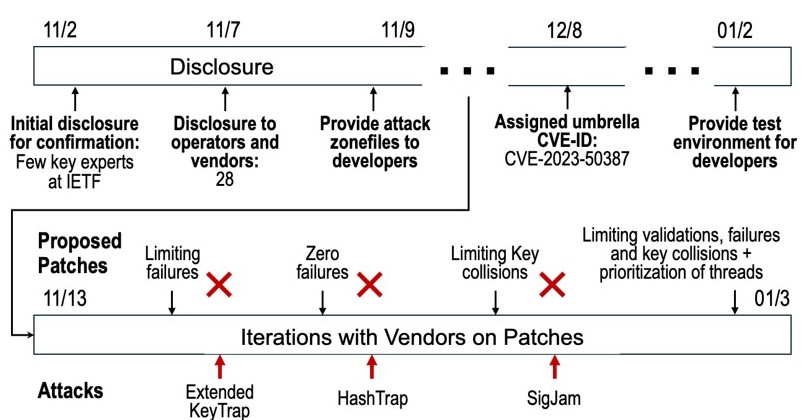}
    \caption{Disclosure and patch/break/fix timeline.}
    \label{fig:timeline}
\end{figure}

The detrimental impact of KeyTrap attacks if exploited in the wild on vulnerable resolvers necessitated patches before the flaws and our attack methodologies become public. We have thus been closely working with the developers of DNS resolvers since November 2, 2023 on developing mitigations against our attacks. We initiated the disclosure process on November 2. 2023, following which a group was formed of 31 participants, consisting of vendors of DNS software, operators of DNS/Cloud/CDN, and IETF experts. The group communicates over a closed DNS OARC channel established for disclosure and mitigations of our attacks. We describe the timeline of disclosure and mitigations in Figure \ref{fig:timeline}. 

The immediate short-term recommendations to mitigate an ongoing attack are to disable DNSSEC or to serve stale data. Serving stale data to improve DNS resiliency was proposed in [RFC8767]. Vendors that decide to implement this should make sure to return stale data from a separate thread, not the one that also does the DNSSEC validation, otherwise the resolvers remain stalled. Disabling DNSSEC validation in resolvers would help remediate an ongoing attack.
However, this would also expose clients and resolvers to threats from DNS cache poisoning. Worse, an adversary could abuse this fallback to insecure mode as means to downgrade the DNSSEC protection. 

We worked with the DNS developers to integrate systematic mitigations into their software. In the following, we describe the succession of proposed patches, showing how we evaluated and circumvented their protection against KeyTrap. The process illustrates the challenges in mitigating such powerful attacks as KeyTrap attacks and variants of it. We also present the first working solution that will be published, in variations, as patches for all major DNS resolvers. The operators of the open DNS resolvers have already deployed patches. The releases of patches for DNS software have been scheduled by the different vendors to be deployed between end of January and beginning of February. It is important to note that these patches all disobey the Internet standard in certain aspects, including the number of validations they are willing to do, to protected against the flaws within the standard. 

\subsection{Patch-Break-Fix DNSSEC}

Agreeing on which patches to deploy required a number of iterations. The developers did not want to make substantial changes, and rather aimed at patches that would mitigate the attacks with minimal changes. This is understandable, since complex patches required more extensive testing over longer time periods to confirm that they do not introduce new flaws, are interoperable with other mechanisms, and do not incur failures in the wild. Nevertheless, developing quick patches turned into a lengthy iterative process, during which the vendors developed patches that we broke, which were subsequently fixed, following with new patches. We illustrate the timeline of the disclosure and the patch-break-fix iterations with the vendors in Figure \ref{fig:timeline}. We next explain the patches and our attacks against them. 

\ignore{limiting signature failures prevents keytrap, limiting collisions prevents hashtrap, limiting key collisions prevents both. limiting key tag collisions does not prevent a variation of keytrap (sigjam).}

\textbf{Limiting failures.} The initial ``immediate'' mitigation was to limit the maximum amount of validation failures per resolution. It was first implemented by Akamai, with a limit of 32 on the number of failed validations, then Bind9, which limited the failures to 0 and Unbound, with a limit of 16 failures. We found the limitation not to be an effective mitigation against our attacks. If each query is limited in the number of failures it is allowed to result in, the failures can be spread across multiple queries. To demonstrate this, we extended the KeyTrap attack (presented in §\ref{sc:keytrap}) so that the signature validations are distributed across multiple queries, such that each query causes the resolver to perform 32 signature validations. Thus instead of creating multiple validations with a single query we sent multiple queries. In a setup with Akamai DNS resolver instance, 150 requests per second cause the CPU to get stalled. This showed that the limit of 32 was not strict enough. 

\begin{figure}[b!]
    \centering
    \includegraphics[width=0.9\columnwidth]{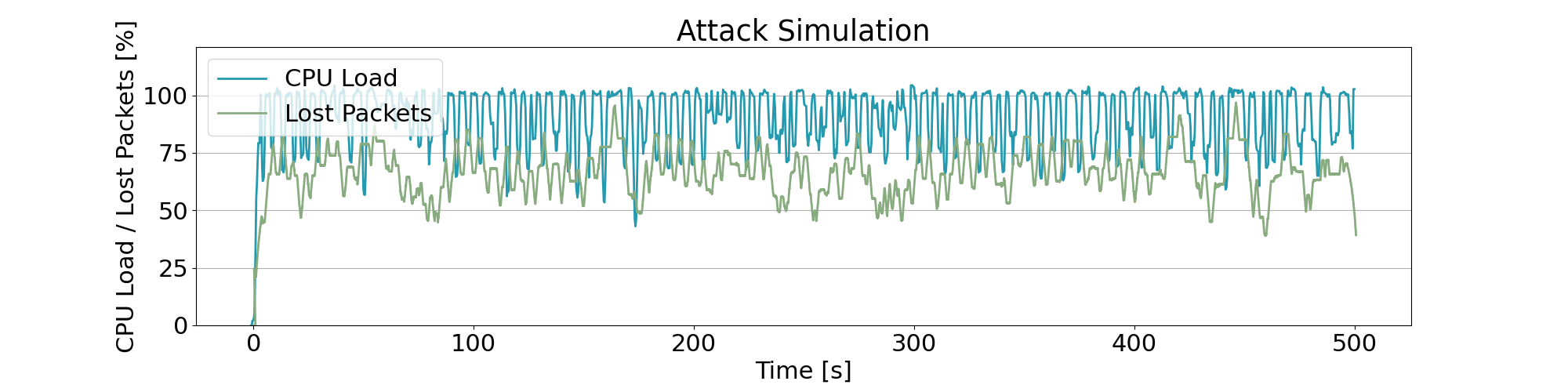}
    \caption{HashTrap attack on patched Bind9 with 10req/s.}
    \label{fig:dshashing}
\end{figure}

\textbf{Zero failures.} The strictest patch on the cryptographic failures was implemented by Bind9, returning SERVFAIL after a single cryptographic failure, hence removing the need to check for collisions at all. Although allowing 0 failed validations prevents the KeyTrap attack, it does not mitigate hash collision attack with HashTrap (§\ref{sc:hashtrap}). 
HashTrap causes the resolver to perform a very large amount of hash calculations. Experimentally, using 10 requests per second, we showed that HashTrap inflicts DoS on the patched instance of Bind9 resolver. The evaluation is plotted in Figure \ref{fig:dshashing}: As can be seen, during the attack against the patched Bind9 instance more than 72\% of benign requests are lost. We observe that most benign requests get dropped. 
This variant of the attack shows that merely limiting the amount of signature validation failures is not a resilient mitigation against our DoS attacks.

\textbf{Limiting key collisions.} 
A patch by Akamai, in addition to limiting the signature validation failures, also limited the key tag collisions in DNS responses to contain at most 4 keys with colliding tags. We find that limiting key tag collisions to 4 will not impact normal operation of the resolver. Using data from the Tranco Top1M domains, we find experimentally that only two zones have colliding DNSKEYS, with no zone using more than two colliding keys.

\begin{figure}[t!]
    \centering
    \includegraphics[width=0.9\columnwidth]{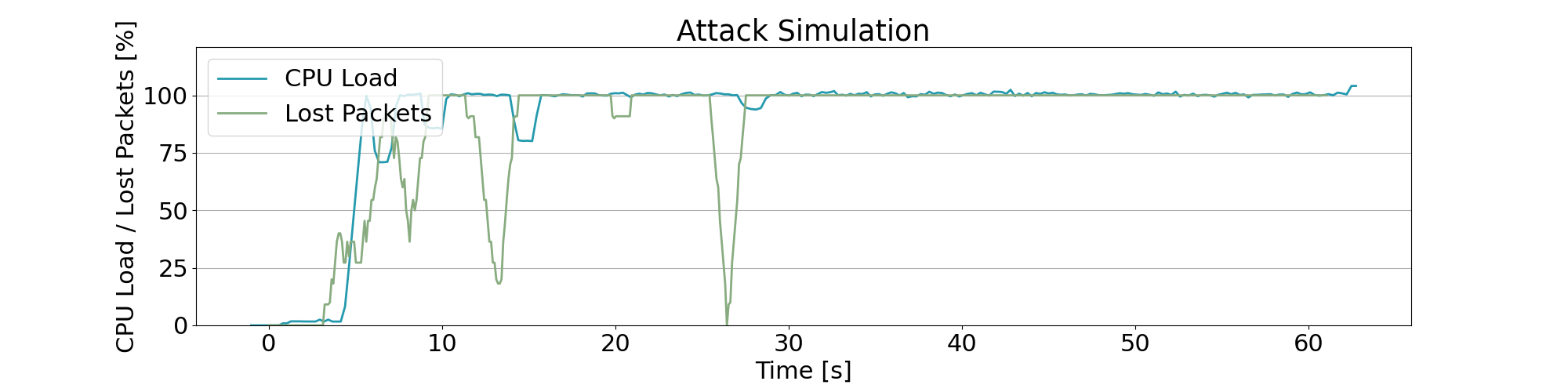}
    \caption{ANY-type derivation of SigJam on Akamai.}
       \vspace{-10pt}
    \label{fig:anytype}
\end{figure}

Limiting key tag collisions proved successful in protecting against HashTrap. 
The combination of both patches was nevertheless still vulnerable to a variant of the SigJam attack (§\ref{sc:sigjam}). The attack works with a single DNSSEC key and many signatures, but requires no signature validation failures, thereby circumventing the protection offered by the patch.
We use ANY type responses, which contain many different record sets, each signed with a different signature.
We can create arbitrary numbers of different record sets, so that on the one hand the number of signatures is maximized, and on the other hand, the response still fits into one DNS packet.
\ccs{We vary over the type number field on an A-type record to create a large number of small, unallocated-type record sets, each covered by an individual, valid signature.
In our tests with standard DNS software, we created DNS responses with 313 different record sets.
Following [RFC6840] §4.2 the resolver \must{} validate the signatures on all the record sets.}
Since all signatures are valid, the resolver does not fail from the imposed limit on validation failures and instead continues the validation until all signatures on the unallocated-type records have been checked. We found this attack to be effective against all patches that limit cryptographic failures. The success of the attack on a patched Akamai is illustrated in Figure \ref{fig:anytype}. In the evaluation, the attacker sends 4 ANY type requests per second, a rate at which the attacker is able to completely DoS the resolver after a few seconds. Running the attack for 60s, we were able to achieve over 90\% lost benign queries. The attack can thus DoS the resolver, circumventing the patch. The attack that exploits ANY type records illustrates that limiting only the cryptographic failures is not a sufficient mitigation against the complexity attacks.

\begin{figure}[b!]
    \centering
    \includegraphics[width=0.9\columnwidth]{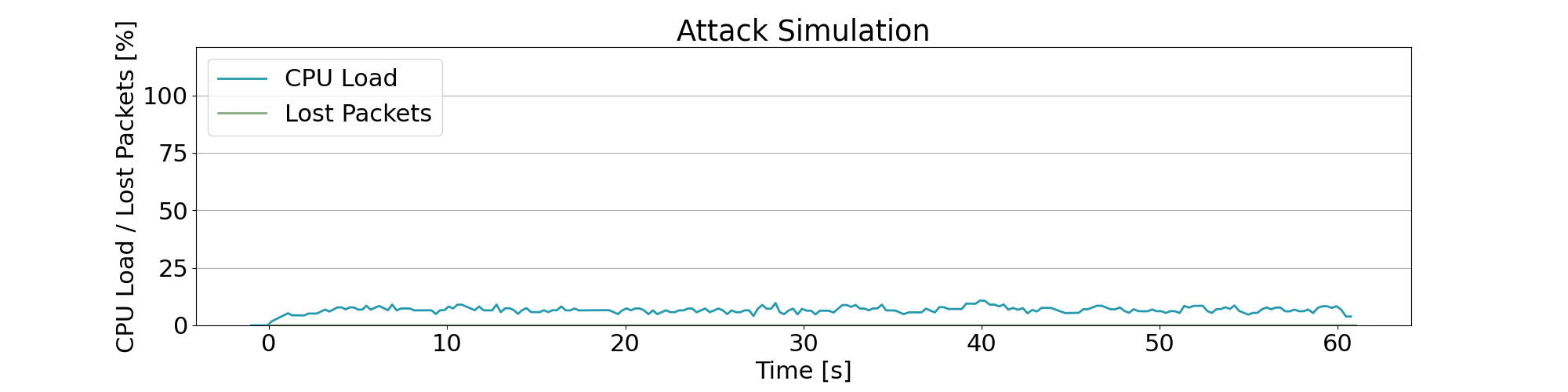}
    \caption{ANY type attack with 10req/s on patched Akamai.}
    \label{fig:cacheservepatch}
\end{figure}

\textbf{Limiting all validations.} 
The first working patch capable of protecting against all variants of our attack was implemented by Akamai. Additionally to limiting key collisions to 4, and limiting cryptographic failures to 16, the patch also limits total validations in ANY requests to 8. Evaluating the efficacy of the patch, we find the patched resolver does not lose any benign request even under attack with > 10 requests per second. Illustrated in Figure \ref{fig:cacheservepatch}, the load on the resolver does not increase to problematic levels under the ANY type attack with 10 req/s, and the resolver does not lose any benign traffic. It thus appears that the patch successfully protects against all variations of KeyTrap attacks. Nevertheless, although these patches prevent packet loss, they still do not fully mitigate the increase in CPU instruction load during the attack.
The reason that the mitigations do not fully prevent the effects of the KeyTrap attacks is rooted in the design philosophy of DNSSEC. \ccs{For example, we find that in a patched Unbound, an attacker request can still displace the processing equivalent of 8 benign requests under full load.} Notice however that we are still closely working with the developers on testing the patches and their performance during attack and during normal operation.

\subsection{Improving Resilience of Architecture}
To understand the impact of the attacks on various DNS functionality, including caching, pending DNS requests or inbound DNS packets from the clients as well as from the nameservers, we perform code analysis and evaluations. Our observations from the analyses can be used to enhance the robustness to failures and attacks of implementations:

{\bf Multi-threading.} Using code analysis and experimental evaluation of the multi-threading architecture of the DNS implementations, we find that load of processes is generally not considered in scheduling new DNS requests, leading to substantial loss of requests even if not all threads of a resolver are busy. Further, we find resolvers do not consider the computational effort of a given request, leading to loss of benign requests, if a single request creates a large load on the resolver. We contribute the architectural recommendation that resolvers should de-prioritize DNS requests that cause substantial computational load, allowing the resolver to still answer benign clients even under attacks. \ccs{This de-prioritization is in line with previous work recommendations on mitigations of complexity attacks, like presented by Atre et al. \cite{atre2022surgeprotector}.}

{\bf OS buffers.} We find that the resolvers generally only deplete the OS UDP buffer after a batch of tasks has been finished. This causes the buffer to fill up when the resolver is busy, leading to lost benign requests. We recommend to adapt the architecture of resolvers to allocate a separate thread for reading from the OS buffer and placing pending requests in a dynamic internal buffer. 

{\bf Thread for cached records.} Further, since many benign queries by users can be answered from cache, additionally allocating a separate thread for answering to cached entries can reduce the impact of stalling of resolution threads.

\subsection{Implementation Challenges}
\begin{figure}[t!]
    \centering
    \includegraphics[width=0.9\columnwidth]{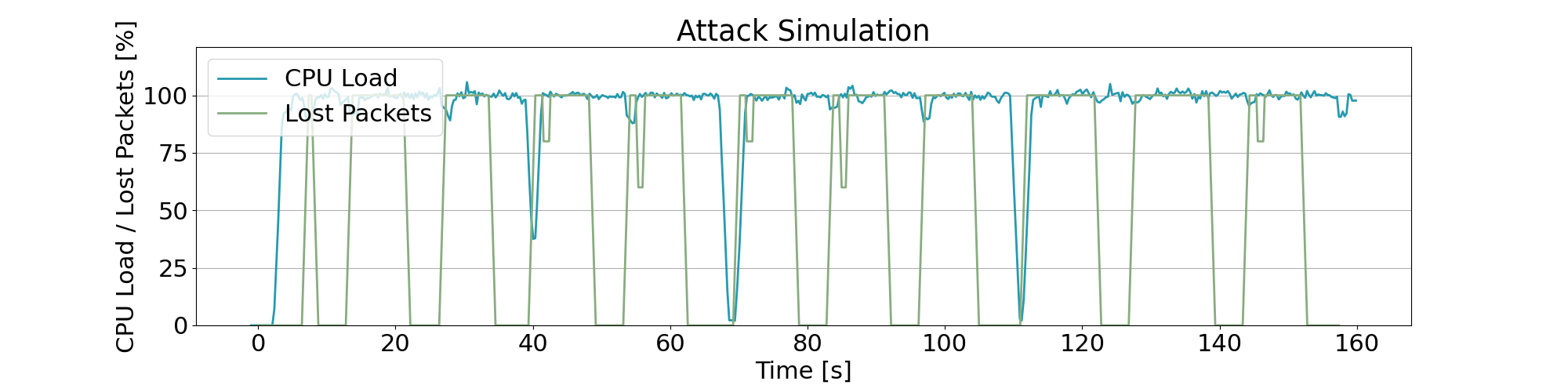}
    \caption{KeyTrap against patched Knot.}
    \label{fig:patched_knot}
\end{figure}

The experience we made working with the developers on designing and evaluating the patches showed that the vulnerabilities we found were challenging to patch. We not only showed that patches could be circumvented with different variants of our attack, but also discovered problems in the implementations themselves. 
We provide here examples from two major implementations: Knot and Bind9. During the evaluations we found that a patch for Knot, that was supposed to limit requests to 32 failed validations per resolution, was not working as intended. While the patch reduced the number of validations resulting from a single attacker request, it did not sufficiently protect against an attacker sending multiple requests in a short time frame. With 10 attacker requests per second, the patched Knot implementation dropped over 60\% of benign queries, as shown in Figure \ref{fig:patched_knot}. We traced the bug to be a broken binding, which the developers fixed in the subsequent iterations of patches.

The second example is a problematic patch in Bind9. While evaluating the patch with 10 requests per second to the patched resolver, we found that after about 70s the resolver would consistently crash, causing 100\% loss of benign queries. This bug was also communicated to developers and fixed in later patches.

\section{Ethical Considerations}\label{sec:ethics}
\ccs{Due to the potentially severe impact of KeyTrap, we limited all our evaluations to a local test setup without testing the KeyTrap attack on any open, publicly accessible resolver. We disclosed the vulnerabilities we found to a closed group of experts over 3 months before they were made public through open-source patches and accompanying posts by the developers. Operators were notified about the imminent important patches with sufficient preparation time and patches were delivered to large operators ahead of time, before the vulnerability became public. We ensured quality of developed patches by continuously working with the developers, improving their patches and closing discovered flaws. From the practical perspective, current patches sufficiently protect against the impact of the attacks.}

\section{Conclusions}\label{sec:conclusions}

Our work revealed a fundamental design problem with DNS and DNSSEC: Strictly applying Postel's Law to the design of DNSSEC introduced a major and devastating vulnerability in virtually all DNS implementations. With just one maliciously crafted DNS packet an attacker could stall almost any resolver, e.g., the most popular one, Bind9, for as long as 16 hours. 

The impact of KeyTrap is far reaching. DNS evolved into a fundamental system in the Internet that underlies a wide range of applications and facilitates new and emerging technologies. Measurements by APNIC\footnote{\url{https://stats.labs.apnic.net/dnssec/XA}} show that in December 2023, $31.47\%$ of the web clients worldwide used DNSSEC-validating resolvers.
Therefore, our KeyTrap attacks have effects not only on DNS but also on any application using it. An unavailability of DNS may not only prevent access to content but risks also disabling security mechanisms, like anti-spam defenses, Public Key Infrastructure (PKI), or even inter-domain routing security like RPKI or rover \cite{karim2019comprehensive,maurer1996modelling,chung2019rpki,gersch2013rover}. 

Since the initial disclosure of the vulnerabilities, we have been working with all major vendors on mitigating the problems in their implementations, but it seems that completely preventing the attacks requires to fundamentally reconsider the underlying design philosophy of DNSSEC, i.e., to revise the DNSSEC standards.

\section*{Acknowledgements}
This work has been co-funded by the German Federal Ministry of Education and Research and the Hessen State Ministry for Higher Education, Research and Arts within their joint support of the National Research Center for Applied Cybersecurity ATHENE and by the Deutsche Forschungsgemeinschaft (DFG, German Research Foundation) SFB~1119.

\bibliographystyle{plain}
\bibliography{ref,bib,valdos}

\begin{thebibliography}{10}

\bibitem{afek2020nxnsattack}
Yehuda Afek, Anat Bremler-Barr, and Lior Shafir.
\newblock $\{$NXNSAttack$\}$: Recursive $\{$DNS$\}$ inefficiencies and
  vulnerabilities.
\newblock In {\em 29th USENIX Security Symposium (USENIX Security 20)}, pages
  631--648, 2020.

\bibitem{afek2023nrdelegationattack}
Yehuda Afek, Anat Bremler-Barr, and Shani Stajnrod.
\newblock $\{$NRDelegationAttack$\}$: Complexity $\{$DDoS$\}$ attack on
  $\{$DNS$\}$ recursive resolvers.
\newblock In {\em 32nd USENIX Security Symposium (USENIX Security 23)}, pages
  3187--3204, 2023.

\bibitem{atre2022surgeprotector}
Nirav Atre, Hugo Sadok, Erica Chiang, Weina Wang, and Justine Sherry.
\newblock Surgeprotector: Mitigating temporal algorithmic complexity attacks
  using adversarial scheduling.
\newblock In {\em Proceedings of the ACM SIGCOMM 2022 Conference}, pages
  723--738, 2022.

\bibitem{bushart2018dns}
Jonas Bushart and Christian Rossow.
\newblock Dns unchained: amplified application-layer dos attacks against dns
  authoritatives.
\newblock In {\em Research in Attacks, Intrusions, and Defenses: 21st
  International Symposium, RAID 2018, Heraklion, Crete, Greece, September
  10-12, 2018, Proceedings 21}, pages 139--160. Springer, 2018.

\bibitem{heartbleed}
Marco Carvalho, Jared DeMott, Richard Ford, and David~A Wheeler.
\newblock Heartbleed 101.
\newblock {\em IEEE security \& privacy}, 12(4):63--67, 2014.

\bibitem{chung2019rpki}
Taejoong Chung, Emile Aben, Tim Bruijnzeels, Balakrishnan Chandrasekaran, David
  Choffnes, Dave Levin, Bruce~M Maggs, Alan Mislove, Roland~van Rijswijk-Deij,
  John Rula, et~al.
\newblock Rpki is coming of age: A longitudinal study of rpki deployment and
  invalid route origins.
\newblock In {\em Proceedings of the Internet Measurement Conference}, pages
  406--419, 2019.

\bibitem{log4j}
Douglas Everson, Long Cheng, and Zhenkai Zhang.
\newblock Log4shell: Redefining the web attack surface.
\newblock In {\em Workshop on Measurements, Attacks, and Defenses for the Web
  (MADWeb) 2022}, 2022.

\bibitem{gersch2013rover}
Joseph Gersch and Dan Massey.
\newblock Rover: Route origin verification using dns.
\newblock In {\em 2013 22nd International Conference on Computer Communication
  and Networks (ICCCN)}, pages 1--9. IEEE, 2013.

\bibitem{heftrig2023downgrading}
Elias Heftrig, Haya Shulman, and Michael Waidner.
\newblock {Downgrading DNSSEC: How to Exploit Crypto Agility for Hijacking
  Signed Zones}.
\newblock In {\em 32nd USENIX Security Symposium (USENIX Security 23)}, pages
  7429--7444, 2023.

\bibitem{kaminsky:dns}
Dan Kaminsky.
\newblock {I}t's the {E}nd of the {C}ache {A}s {W}e {K}now {I}t.
\newblock In {\em Black Hat conference}, August 2008.
\newblock
  \url{http://www.blackhat.com/presentations/bh-jp-08/bh-jp-08-Kaminsky/BlackHat-Japan-08-Kaminsky-DNS08-BlackOps.pdf}.

\bibitem{karim2019comprehensive}
Asif Karim, Sami Azam, Bharanidharan Shanmugam, Krishnan Kannoorpatti, and
  Mamoun Alazab.
\newblock A comprehensive survey for intelligent spam email detection.
\newblock {\em IEEE Access}, 7:168261--168295, 2019.

\bibitem{kuhrer2014exit}
Marc K{\"u}hrer, Thomas Hupperich, Christian Rossow, and Thorsten Holz.
\newblock Exit from hell? reducing the impact of
  $\{$Amplification$\}$$\{$DDoS$\}$ attacks.
\newblock In {\em 23rd USENIX security symposium (USENIX security 14)}, pages
  111--125, 2014.

\bibitem{maurer1996modelling}
Ueli Maurer.
\newblock Modelling a public-key infrastructure.
\newblock In {\em Computer Security—ESORICS 96: 4th European Symposium on
  Research in Computer Security Rome, Italy, September 25--27, 1996 Proceedings
  4}, pages 325--350. Springer, 1996.

\bibitem{DBLP:conf/imc/MouraCHWH20}
Giovane C.~M. Moura, Sebastian Castro, Wes Hardaker, Maarten Wullink, and
  Cristian Hesselman.
\newblock Clouding up the internet: how centralized is {DNS} traffic becoming?
\newblock In {\em Internet Measurement Conference}, pages 42--49. {ACM}, 2020.

\bibitem{van2016performance}
Roland van Rijswijk-Deij, Kaspar Hageman, Anna Sperotto, and Aiko Pras.
\newblock The performance impact of elliptic curve cryptography on dnssec
  validation.
\newblock {\em IEEE/ACM transactions on networking}, 25(2):738--750, 2016.

\bibitem{van2014dnssec}
Roland van Rijswijk-Deij, Anna Sperotto, and Aiko Pras.
\newblock Dnssec and its potential for ddos attacks: a comprehensive
  measurement study.
\newblock In {\em Proceedings of the 2014 Conference on Internet Measurement
  Conference}, pages 449--460, 2014.

\end{thebibliography}

\balance
\appendix

\end{document}